\documentclass[review]{elsarticle}

\journal{Information Sciences}
\usepackage[pagewise]{lineno}
\usepackage{array,multirow}
\usepackage{caption}
\usepackage{graphicx}
\usepackage{subfig}
\usepackage{amsmath}
\usepackage{geometry}
\usepackage[numbers]{natbib}
\newcolumntype{P}[1]{>{\centering\arraybackslash}p{#1}}
\newcolumntype{M}[1]{>{\centering\arraybackslash}m{#1}}
\usepackage{hyperref}
\hypersetup{colorlinks=true,linkcolor=blue,anchorcolor=blue,citecolor=blue,filecolor=blue,urlcolor=blue,bookmarksnumbered=true,pdfview=FitB,linktocpage=true,hypertexnames=false,pdftex}

\usepackage{listings,color}
\usepackage[pscoord]{eso-pic}
\newcommand{\placetextbox}[3]{
	\setbox0=\hbox{#3}
	\AddToShipoutPictureFG{ \put(\LenToUnit{#1\paperwidth},\LenToUnit{#2\paperheight}){\vtop{{\null}\makebox[0pt][c]{#3}}}
	}
}
\placetextbox{.5}{0.997}
{\fontsize{6}{6}\selectfont
	This article has been accepted and published in the journal ``Information Sciences". Content may change prior to final publication.  
}
\placetextbox{.5}{0.988}
{\fontsize{6}{6}\selectfont
	DOI: https://doi.org/10.1016/j.ins.2019.04.019, 0020-0255/ \textcopyright \: 2019 Elsevier Inc.All rights reserved.
}

\placetextbox{.5}{0.035}
{\fontsize{6}{6}\selectfont
	Copyright \textcopyright \: 2019 Elsevier Inc. All rights reserved. This manuscript version is made available under the license http://creativecommons.org/licenses/by-nc-nd/4.0/
}

\begin{document}

\begin{frontmatter}

\title{A New Approach for Query Expansion using Wikipedia and WordNet}

\author[]{Hiteshwar Kumar Azad \corref{myca}}
\ead{hiteshwar.cse15@nitp.ac.in}
\cortext[myca]{Corresponding Author}%\fntext[myfootnote]{Since 1880.}

\author[]{Akshay Deepak}
\ead{akshayd@nitp.ac.in}

\address{Dept. of Computer Science \& Engineering\\
	National Institute of Technology Patna, India}

\begin{abstract}
Query expansion (QE) is a well-known technique used to enhance the effectiveness of information retrieval. QE reformulates the initial query by adding similar terms that help in retrieving more relevant results. Several approaches have been proposed in literature producing quite favorable results, but they are not evenly favorable for all types of queries (individual and phrase queries). One of the main reasons for this is the use of the same kind of data sources and weighting scheme while expanding both the individual and the phrase query terms. As a result, the holistic relationship among the query terms is not well captured or scored. To address this issue, we have presented a new approach for QE using Wikipedia and WordNet as data sources. Specifically, Wikipedia gives rich expansion terms for phrase terms, while  WordNet does the same for individual terms. We have also proposed novel weighting schemes for expansion terms: \emph{in-link score} (for terms extracted from Wikipedia) and a tf-idf based scheme (for terms extracted from WordNet). In the proposed Wikipedia-WordNet-based QE technique (WWQE), we weigh the expansion terms twice:  first, they are scored by the weighting scheme individually, and then, the weighting scheme scores the selected expansion terms concerning the entire query using correlation score. The proposed approach gains improvements of 24\% on the MAP score and 48\% on the GMAP score over unexpanded queries on the FIRE dataset. Experimental results achieve a significant improvement over individual expansion and other related state-of-the-art approaches. We also analyzed the effect on retrieval effectiveness of the proposed technique by varying the number of expansion terms. 
\end{abstract}

\begin{keyword}
\texttt{Query Expansion \sep Information Retrieval \sep WordNet \sep Wikipedia}
\end{keyword}

\end{frontmatter}

\section{Introduction}\label{intro}
The Web is the largest information source available on the planet and it's growing day by day \cite{merigo2018fifty}. According to a recent survey \cite{chakraborty2014analysis} in Computerworld magazine, approximately 70--80 percent of all data available to enterprises/organizations is unstructured information, i.e., information that either is not organized in a pre-defined manner or does not have a pre-defined data model. This makes information processing a big challenge and creates a vocabulary gap between user queries and indexed documents. It is common for a user's query $Q$ and its set of relevant documents $D$ to use different vocabulary and language styles while referring to the same concept. For example, terms `buy' and `purchase' have the same meaning; however, only one of these can be present in the documents' index while the other one can be the user's query term.  This makes it difficult to retrieve the information actually wanted by the user \cite{ji2019query,liu2018perceptual}. An effective strategy to fill this gap is to use the Query Expansion (QE) technique, which enhances the retrieval effectiveness by adding expansion terms to the initial query. Selection of the expansion terms plays a crucial role in QE because only a small subset of the expanded terms are actually relevant to the query \cite{liu2017multi}. In this sense, the approach for selection of expansion terms is equally important in comparison to what we do further with the expanded terms in order to retrieve the desired information.  QE has a long research history in information retrieval (IR) \cite{maron1960relevance,rocchio1971relevance}. It has the potential to enhance IR's effectiveness by adding relevant terms that can help to discriminate the relevant documents from irrelevant ones. The source of expansion terms plays a significant role in QE. A variety of sources have been researched for extracting the expansion terms, e.g., the entire target document collection \cite{bai2005query}, feedback documents (few top ranked documents are retrieved in response to the initial query) \cite{li2007improving}, or external knowledge resources \cite{dang2010query}.

References \cite{carpineto2012survey,azad2017query} provide comprehensive surveys on  data sources used for QE. Broadly, such sources can be classified into four categories: (i) documents used in the retrieval process \cite{bai2005query} (e.g., corpus), (ii) hand-built knowledge resources \cite{pal2014improving} (e.g., WordNet\footnote{https://wordnet.princeton.edu/}, ConceptNet\footnote{http://conceptnet5.media.mit.edu/}, thesaurus, ontologies),  (iii) external text collections and resources \cite{dang2010query} (e.g., Web, Wikipedia), and (iv) hybrid data sources \cite{dalton2014entity}.

In corpus-based sources, a corpus is prepared that contains a cluster of terms for each possible query term. During expansion, the corresponding cluster is used as the set of expanded terms (e.g., \cite{bai2005query,dong2018clustering}). However, corpus-based sources fail to establish a relationship between a word in the corpus and related words used in different communities, e.g., ``senior citizen" and ``elderly". 

Hand-built knowledge-resources-based QE extracts knowledge from textual hand-built data sources such as dictionaries, thesauruses, ontologies, and the LOD cloud (e.g., \cite{voorhees1994query,zhang2009concept,pal2014improving}). Thesaurus-based QE can be either automatic or hand-built. One of the most famous hand-built thesauruses is WordNet. While it significantly improves the retrieval effectiveness of badly constructed queries,  it does not show a lot of improvement for  well-formulated user queries. Primarily, there are three limitations of hand-built knowledge resources: they are commonly domain-specific, they usually do not contain proper nouns, and they have to be kept up to date.

External text collections and resources such as web, Wikipedia, query logs and anchor texts are the most common and effective data sources for QE \cite{dang2010query}. In such cases, QE approaches show overall better results in comparison to the other previously discussed data sources.

Hybrid data sources are a combination of two or more data sources. For example, reference \cite{collins2005query} uses WordNet,  an external corpus,  and the top retrieved documents as data sources for QE. Some of the other research works for query processing based on hybrid resources are \cite{dalton2014entity,li2018skqai,fan2018handling}.

Among the above data sources, Wikipedia and WordNet are popular choices for semantic enrichment of the initial query \cite{voorhees1994query,pal2014improving,almasri2013wikipedia}. They are also two of the most widely used knowledge resources in natural language processing. Wikipedia is the largest encyclopedia describing entities. WordNet is a large lexicon database of words in the English language. An entity is described by Wikipedia through a web article that contains detailed related information about the entity. Each such web article describes only one entity. The information present in the article has important keywords that can prove very useful as expansion terms for queries based on the entity being described by the article. On the other hand, WordNet consists of a graph of synsets that are collections of synonymous words linked by a number of useful properties. WordNet also provides a precise and attentively assembled hierarchy of  useful concepts. These features make WordNet an ideal knowledge resource for QE.  

Many of the articles \cite{voorhees1994query,liu2004effective,pal2014improving,almasri2013wikipedia} have used Wikipedia and WordNet separately with promising results. However, they don't produce consistent results for different types of queries (individual and phrase queries). 

This article proposes a novel technique named \textit{Wikipedia-WordNet based QE technique} (WWQE) for query expansion that combines Wikipedia and WordNet data sources to improve retrieval effectiveness.  We have also proposed novel schemes for weighting expanded terms: \emph{in-link score} (for terms extracted from Wikipedia) and a tf-idf-based scheme (for terms extracted from WordNet).  Experimental results show that the proposed WWQE technique produces consistently better results for all kinds of queries (individual and phrase queries) when compared with query expansion based on the two data sources individually and other related state-of-the-art approaches. We also analyzed the effect on retrieval effectiveness of the proposed technique by varying the number of expansion terms.  The experiments were carried out on the FIRE\footnote{http://fire.irsi.res.in/fire/static/data} dataset using popular weighting models and evaluation metrics.

\subsection{Contributions}
The contributions of this paper are as follows:
\begin{itemize}
	\item \emph{Data Sources.} A novel technique for query expansion named \textit{Wikipedia-WordNet-based QE technique} (WWQE) is proposed that combines Wikipedia and WordNet as data sources. To the best of our knowledge, these two data sources have not been used together for QE.  
	\item \emph{Term Selection.} The proposed WWQE technique employs a two-level strategy to select terms from WordNet based on a proposed tf-idf-based weighting scheme. For fetching expansion terms from Wikipedia pages of the query terms, the proposed technique uses a novel weighting scheme based on out-links and in-links called \emph{in-link score}. 
	%First, it fetches synsets of the initial query terms. Then, it extracts sysnets of these synsets. 
	
	\item \emph{Phrase Selection.} To deal with the user's initial query, the proposed WWQE technique selects both phrases and individual words. A phrase usually offers a richer context and has less ambiguity. Hence, the expansion terms retrieved in response to phrases are more effective than expansion terms in response to the non-phrase terms from the initial query. 
	\item \emph{Weighting Method.} For weighting candidate expansion terms obtained from Wikipedia, the proposed WWQE technique uses a novel  weighting scheme based on out-links and in-links, and correlation score. For terms obtained from WordNet, it uses a novel tf-idf and correlation-score-based weighting scheme. 
	\item Experiments were conducted on Forum for Information Retrieval Evaluation (FIRE) collections. They produced improved results on popular metrics such as  MAP (Mean Average Precision), GM\_MAP (Geometric Mean Average Precision), P@10 (Precision at Top 10 Ranks), P@20, P@30, bpref (binary preference) and overall recall. The comparison was made with results obtained on individual data sources (i.e., Wikipedia and WordNet) and other state-of-the-art approaches. We also analyzed the effect on retrieval effectiveness of the proposed technique by varying the number of expansion terms. 
\end{itemize}

\subsection{Organization} 
The remainder of the article is organized as follows. Section \ref{Related Work} discusses related work. Section \ref{Our Approach} describes the proposed approach. Experimental Setup, dataset and evaluation matrices are discussed in Section \ref{Experimental Setup}. Section \ref{Experimental Results} discusses the experimental results. Finally, we conclude the study in Section \ref{Conclusion}.

\section{Related Work}
\label{Related Work}
There is a wealth of literature on query expansion in the area of information retrieval (IR). In the 1960s, Moron and Kuhns \cite{maron1960relevance} were the first researchers who applied QE  for literature indexing and searching in a mechanized library system. In 1971, Rocchio \cite{rocchio1971relevance} put QE under the spotlight through the ``relevance feedback method'' and its characterization in a vector space model. This method is still used in its original and modified forms in automatic query expansion (AQE). Rocchio's work was further extended and applied in techniques such as collection-based term co-occurrence, cluster-based information retrieval, comparative analysis of term distribution, and automatic text processing \cite{stein2019analysis,lee2019memetic}.

Recently,  QE has been focused on because a lot of researchers are using QE techniques for working on personalized social bookmarking services \cite{zhou2018iterative}, Question Answering over Linked Data (QALD)\footnote{http://qald.sebastianwalter.org/}, Text Retrieval Conference (TREC)\footnote{http://trec.nist.gov/}, and Forum for Information Retrieval Evaluation (FIRE)\footnote{http://fire.irsi.res.in/} collections. They are also used heavily in web, desktop and email searches \cite{pal2015exploring,pedrycz2018computational,nakamura2019anatomy}. Many platforms provide a QE facility to end users, which can be turned on or off, e.g., WordNet\footnote{https://wordnet.princeton.edu/}, ConceptNet\footnote{http://conceptnet5.media.mit.edu/}, Lucene\footnote{http://lucene.apache.org/}, Google Enterprise\footnote{https://enterprise.google.com/search/} and MySQL. 
Some surveys have previously been done on QE techniques. In 2007, Bhogal et al. \cite{bhogal2007review}  reviewed QE techniques using ontologies, which are domain-specific. Such techniques have also been described in book \cite{Manning:2008:IIR:1394399}. Carpineto and Romano \cite{carpineto2012survey} reviewed major QE techniques, data sources and features in an information retrieval system. In this paper we propose an AQE technique based on WordNet and Wikipedia, which are currently highly influential data sources. These two sources are described next. 

\subsection{Use of WordNet as a Data Source for QE}
WordNet is one of the most popular hand-built thesauruses and has been significantly used for QE and word-sense disambiguation (WSD). Here, our focus is on the use of WordNet for query expansion. There are many issues that need to be addressed when using WordNet as a data source, such as:
\begin{itemize}
	\item When a query term appears in multiple synsets, which synset(s) should be considered for query expansion?
	\item Can only the synsets of a query term have meanings similar to the query term, or can synsets of these synsets also have meanings similar to the query term and hence should they also be considered as potential expansion terms? 
	\item When considering a synset of a query term, should only synonyms be considered or should other relations (i.e., hypernyms, hyponyms, holonyms, meronyms etc.) also be looked at? Further, when considering terms under a given relation, which terms should be selected? 
	
\end{itemize}
In earlier works, a number of researchers have explored these issues. Reference \cite{voorhees1994query} added manually selected WordNet synsets for QE, but unfortunately no significant improvements were obtained. Reference \cite{smeaton1995trec} used synonyms of the initial query and assigns half the weight. Reference \cite{liu2004effective} used word sense to add synonyms, hyponyms and terms' WordNet glosses to expand the query. Their experiments yielded significant improvements on TREC datasets. Reference \cite{zhang2009concept} used sense disambiguation of query terms to add synonyms for QE. During experimental evaluation, in response to the user's initial query, reference \cite{zhang2009concept}'s method produced an improvement of around 7\% in P@10 value over the CACM collection. Reference \cite{fang2008re} used a set of Candidate Expansion Terms (CET) that included all the terms from all the synsets where the query terms exist. Basically, a CET is chosen based on the vocabulary overlap between its glosses and the glosses of query terms. 

Recently, reference \cite{pal2014improving} used semantic relations from the WordNet. The authors proposed a novel query expansion technique where Candidate Expansion Terms (CET) are selected from a set of pseudo-relevant documents. The usefulness of these terms is determined by considering multiple sources of information. The semantic relation between the expanded terms and the query terms is determined using WordNet. On the TREC collection, their method showed significant improvement in IR over the user's unexpanded queries. Reference \cite{lemos2014thesaurus} presents an automatic query expansion (AQE) approach that uses word relations to increase the chances of finding relevant code. As data source for query expansion, it uses a thesaurus containing only software-related word relations along with WordNet. More recently, reference \cite{lu2015query} used WordNet for effective code search, where it was used to generate synonyms. These synonyms were used as query expansion terms. During experimental evaluation, their approach showed improvements in precision and recall values of 5\% and 8\% respectively.

In almost all the aforementioned studies, CETs were taken from WordNet as synsets of initial queries. In contrast, we selected CETs from not only the synsets of the initial query, but also synsets of these synsets. We then assigned weights to the synonyms level-wise.

\subsection{ Use of Wikipedia as Data Source for QE}
 
Wikipedia is freely available and is the largest multilingual online encyclopedia on the web, where articles are regularly updated and new articles are added by a large number of web users. The exponential growth and reliability of Wikipedia makes it an ideal knowledge resource for information retrieval.

Recently, Wikipedia has been used widely for QE and a number of studies have reported  significant improvements in IR over TREC and Cultural Heritage in CLEF (CHiC) datasets (e.g., \cite{li2007improving,elsas2008retrieval,almasri2013wikipedia}). Reference \cite{li2007improving} performed an investigation using Wikipedia and retrieved all articles corresponding to the original query as a source of expansion terms for pseudo relevance feedback. It observed that for a particular query where the usual pseudo relevance feedback fails to improve the query, Wikipedia-based pseudo relevance feedback improves it significantly. Reference \cite{elsas2008retrieval} used link-based QE on Wikipedia and focused on anchor text. It also proposed a phrase-scoring function. Reference \cite{xu2009query} utilized Wikipedia to categorize the original query into three types: (1) ambiguous queries (queries with terms having more than one potential meaning), (2) entity queries (queries having a specific meaning that covers a narrow topic),  and (3) broader queries (queries having neither an ambiguous nor specific meaning). They consolidated the expansion terms into the original query and evaluated these techniques using language modelling IR. Reference \cite{almasri2013wikipedia} uses Wikipedia for semantic enrichment of short queries based on in-link and out-link articles. Reference \cite{dalton2014entity} proposes the Entity Query Feature Expansion (EQFE) technique. It uses data sources such as Wikipedia and Freebase to expand the initial query with features from entities and their links to knowledge bases (Wikipedia and Freebase). It also uses structured attributes and the text of the knowledge bases for query expansion. The main motive for linking entities to knowledge bases is to improve the understanding and representation of text documents and queries.

Our proposed WWQE method differs from the above-mentioned expansion methods in three ways:
\begin{enumerate}
	\item Our method uses both Wikipedia and WordNet for query expansion, whereas the above-discussed methods either use only one of these sources or some other sources. 
	\item For extracting expansion terms from WordNet, our method employs a novel two-level approach where synsets of the query term as well as the synsets of these synsets are selected. Here, terms are scored by a newly proposed tf-idf-based weighting scheme.
	\item For extracting expansion terms from Wikipedia, terms are selected on the basis of a novel scheme called `in-link score', which is based on in-links and out-links of Wikipedia articles. 
\end{enumerate}

\section{Our Approach}
\label{Our Approach}
We present a new way of QE combining Wikipedia and WordNet as data sources. In order to expand the user's query, first, we pre-process the initial query to extract the phrases and individual terms. These phrases and individual terms are used to obtain similar terms from Wikipedia and WordNet. After this, we assign the similarity score to these terms using the proposed weighting score. This is done  separately for terms from Wikipedia and WordNet. Then, the top $n$ terms from Wikipedia and WordNet are selected as expansion terms. These expansion terms are re-weighted based on the correlation score. Finally, we select the top $m$ terms as the final set of expansion terms.

\begin{figure}[!h]
	\centering 
	\includegraphics[width=13cm, height=14cm]{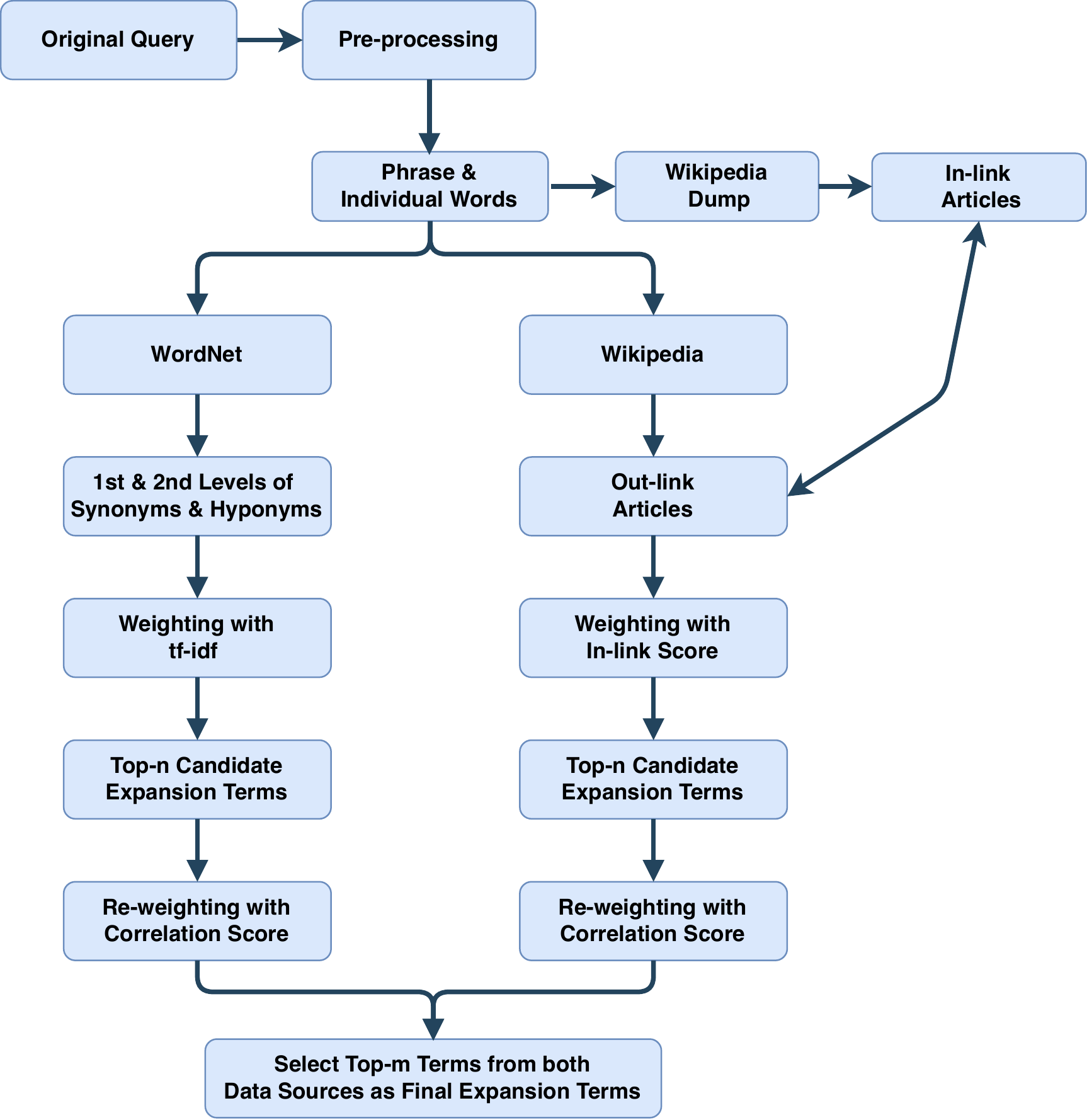}  
	%\captionsetup{justification=centering}    
	\caption{Steps involved in the proposed approach}
	\label{art} 
\end{figure}

The proposed approach consists of four main steps: (i) Pre-processing of initial query, (ii) QE using Wikipedia, (iii) QE using WordNet, and (iv) Re-weighting of expanded terms. Figure \ref{art} summarizes these steps.
\subsection{Pre-processing of Initial Query}
In the Preprocessing step, Brill's tagger is used to lemmatize each query and assign a Part Of Speech (POS) to each word in the query. The POS tagging is done on the queries and the POS information is used to recognize the phrase and individual words. These phrases and individual words are used in the subsequent steps of QE. Many researchers agree that instead of considering the term-to-term relationship, dealing with the query in terms of phrases gives better results \cite{al2014wikipedia}. Phrases usually offer richer context and have less ambiguity. Hence, the documents retrieved in response to the phrases from the initial query have more importance than the  documents retrieved in response to the non-phrase words from the initial query. A phrase usually has a specific meaning that goes beyond the  cumulative  meaning of the individual component words. Therefore, we give more priority to phrases in the query than the individual words when finding expansion terms from Wikipedia and WordNet. 

For example, consider the following query (Query ID- 126) from the FIRE dataset to demonstrate our pre-processing approach:
\begin{lstlisting}[language=xml]
<top>
<num>126</num>
<title>Swine flu vaccine</title>
<desc>Indigenous vaccine made in India for swine flu prevention</desc>
<narr>Relevant documents should contain information related 
to making indigenous swine flu vaccines in India, the vaccines
use on humans and animals, arrangements that are in place to 
prevent scarcity / unavailability of the vaccine, and the
vaccines role in saving lives.</narr>
</top>
\end{lstlisting}

Multiple such queries in the standard SGML format are present in the query file of the FIRE dataset. For extracting the root query, we extract the title from each query and tag it using the Stanford POS tagger library.
For example, after POS tagging, the title of the above query looks like:\\  \emph{Swine\_NN flu\_NN vaccine\_NN}.\\
For extracting phrases, we have only considered nouns, adjectives, and verbs as the words of interest. We consider a phrase to have been identified whenever two or more consecutive nouns, adjectives, or verbs, or words with the cardinal number are found. Based on this, we get the following individual terms and phrases from the above query:\\  
\emph{Swine}\\
\emph{flu}\\
\emph{Swine flu}\\
\emph{vaccine}\\
\emph{flu vaccine}\\
\emph{Swine flu vaccine}

\subsection{QE using Wikipedia}
\begin{figure}[h]
	\centering 
	\includegraphics[width=12cm, height=3.5cm]{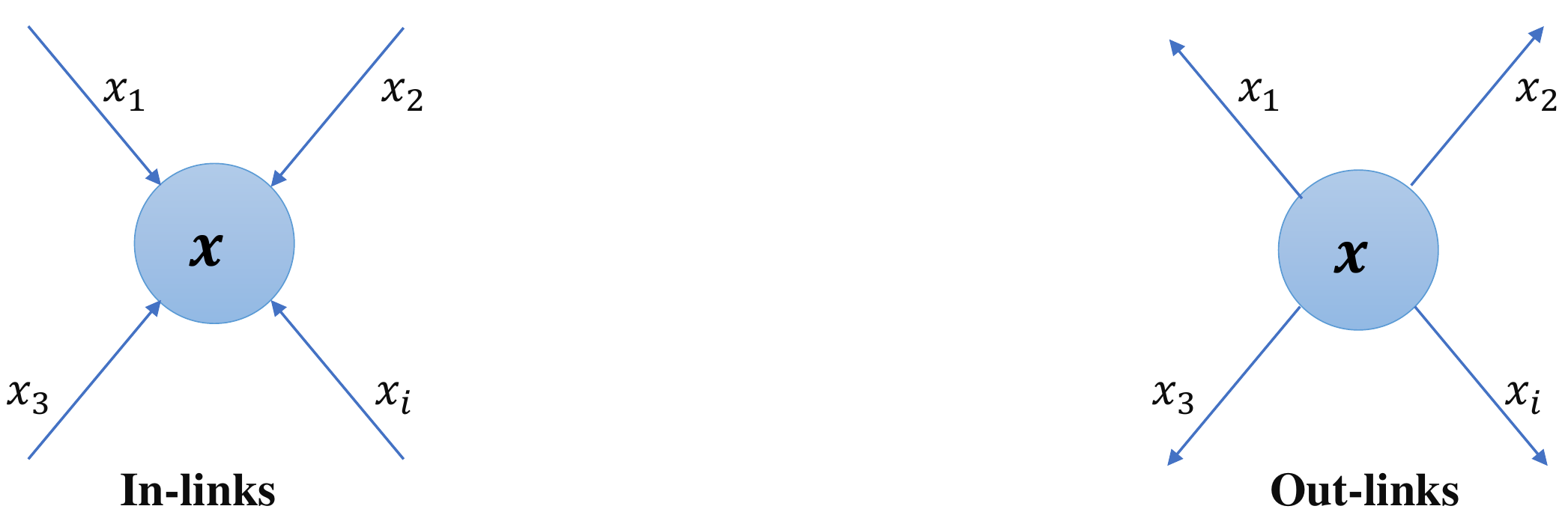}  
	%\captionsetup{justification=centering}    
	\caption{In-links and out-links structure of Wikipedia}
	\label{fig:In-links Out-links structure} 
\end{figure}

After pre-processing the initial query, we consider individual words and phrases as keywords to expand the initial query using Wikipedia. To select CETs from Wikipedia, we mainly focus on Wikipedia titles, in-links and out-links. Before going into further details, we first discuss our Wikipedia representation.
\\ \\
{ \bf \emph{ Wikipedia Representation}}
\\ 
Wikipedia is an ideal information source for QE and can be represented as a directed graph $G(A,L)$, where $A$ and $L$ indicate articles and links respectively. Each article $x \in A$ effectively summarizes its entity, i.e., the title of $x$: $title (x)$, and provides links to the user to browse other related articles. In our work, we consider two types of links: in-links and out-links.
\\ 
{ \bf \emph{ In-links }} of $x$, denoted $I(x)$, is the set of articles that point to the article $x$. It can be defined as 
\begin{equation}
I(x) = \{x_i \mid (x_i, x) \in L\}
\end{equation}
For example, assume we have an article titled ``Computer Science". The in-links to this article  will be all the titles in Wikipedia that hyperlink to the article titled ``Computer Science" in their main text or body. 
\\ 
{ \bf \emph{ Out-links }} of $x$, denoted $O(x)$ is the set of articles that $x$ point to. It can be defined as 
\begin{equation}
O(x) = \{x_i \mid (x, x_i) \in L\}
\end{equation} 
For example, again consider the article titled ``Computer Science". The out-links refer to all the hyperlinks within the body of the Wikipedia page of the article titled ``Computer Science" (i.e., $https://en.wikipedia.org/wiki/Computer\_Science$). The in-links and out-links have been diagrammatically demonstrated in Fig.  \ref{fig:In-links Out-links structure}.

In addition to the article pages, Wikipedia contains ``redirect" pages that provide an alternative way to reach the target article for abbreviated query terms. For example, the query ``ISRO" redirects to the article ``Indian Space Research Organisation" and ``UK" redirects to ``United Kingdom".

In our proposed WWQE approach, the following steps are taken for QE using Wikipedia.
\begin{itemize}
	%	\item \emph{Partitioning of the Wikipedia Dump.}
	\item \emph{Extraction of in-links.}
	\item \emph{Extraction of out-links.} 
	\item \emph{Assignment of the in-link score to expansion terms.}
	\item \emph{Selection of top n terms as expansion terms.}
	\item \emph{Re-weighting of expansion terms.}
\end{itemize}

\subsubsection{Extraction of in-links}

\begin{figure}[h]
	\centering 
	\includegraphics[width=13cm, height=5.2cm]{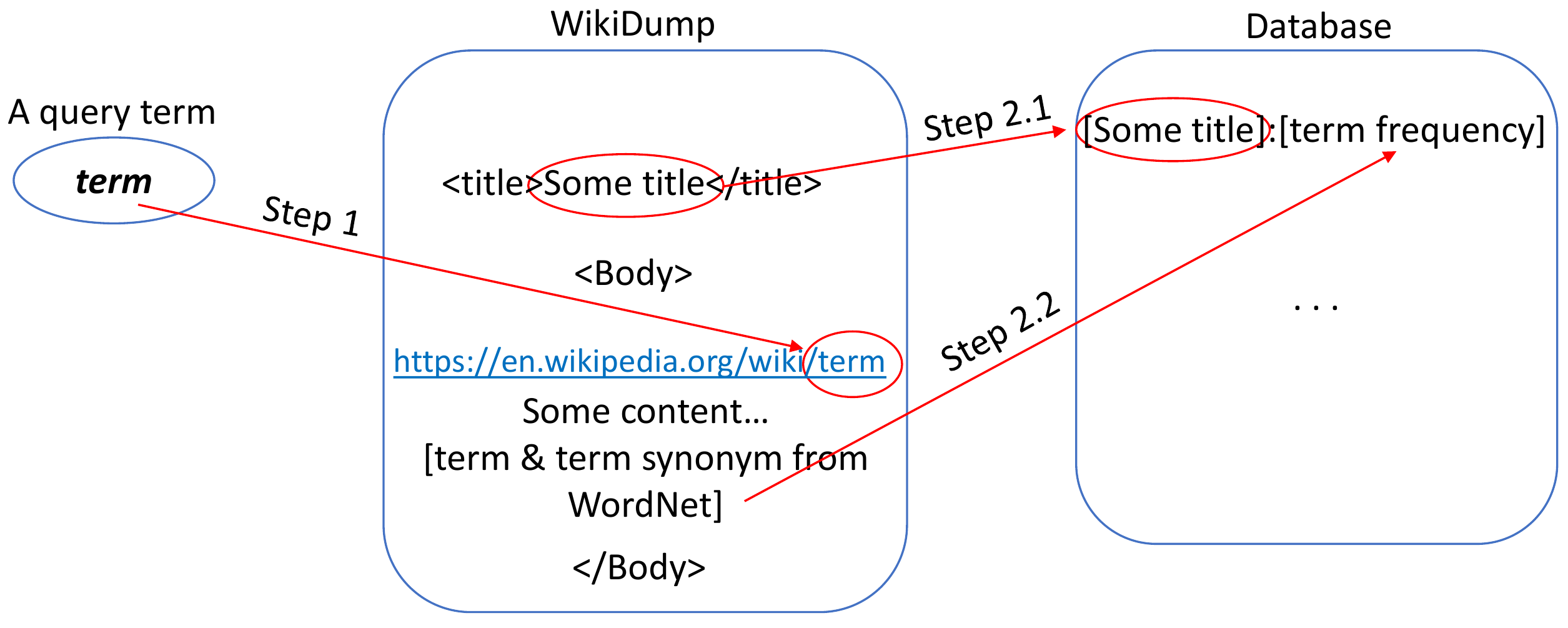}  
	%\captionsetup{justification=centering}    
	\caption{In-links Extraction}
	\label{fig:2} 
\end{figure}

This step involves two sub-steps: (i) extraction of in-links and (ii) computation of term frequency ($tf$) of the initial query terms. The in-links of an initial query term consist of the titles of all those Wikipedia articles that contain a hyperlink to the given query term in their main text or body. $tf$ of an initial query term is the term frequency of the initial query term and its synonyms obtained from WordNet in the in-link articles (see Fig.\ref{fig:2}). For example, if the initial query term is ``Bird", and ``Wings" is one of its in-links, then $tf$ of ``Bird" in the article ``Wings" is the term frequency of the word ``Bird" and its synonyms obtained from WordNet in the article ``Wings".

\subsubsection{Extraction of Out-links}

\begin{figure}[h]
	\centering 
	\includegraphics[width=13cm, height=5.2cm]{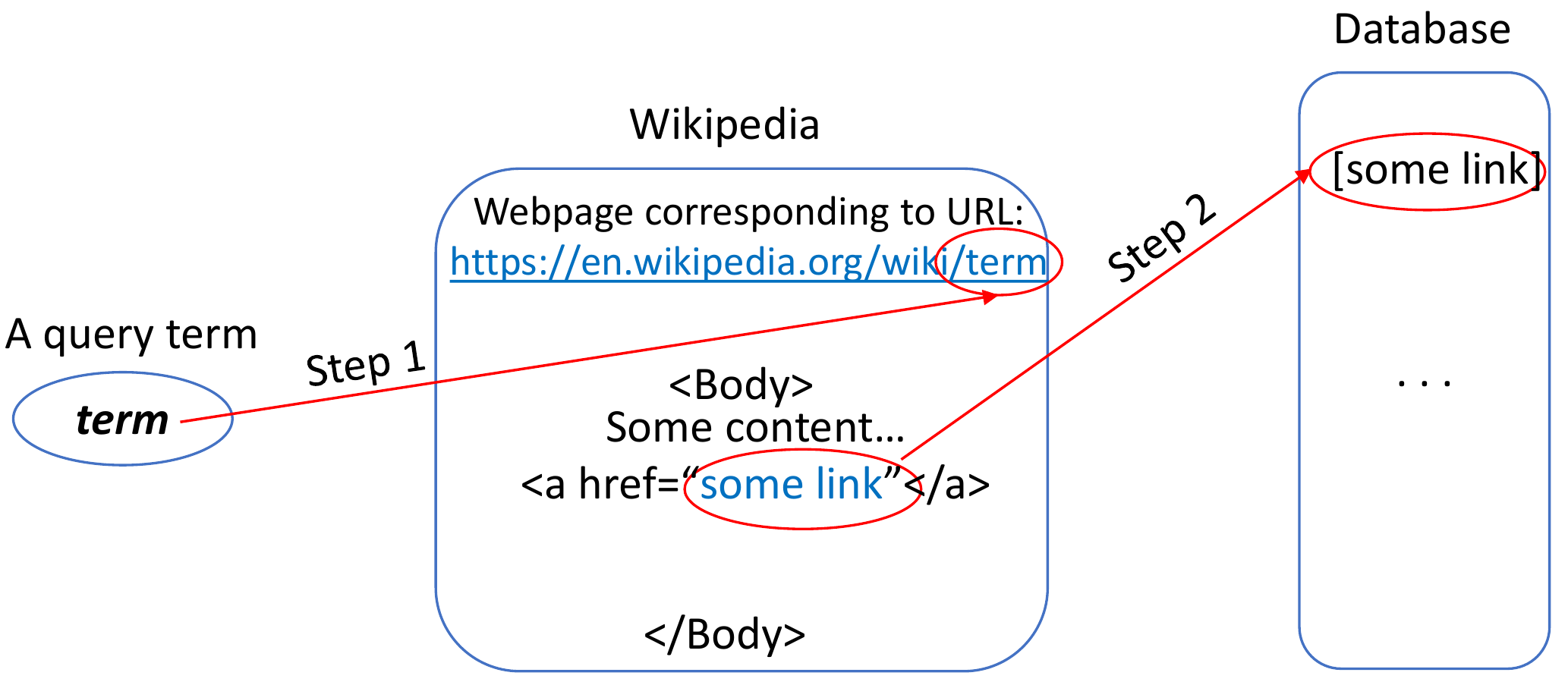}  
	%\captionsetup{justification=centering}    
	\caption{Out-links Extraction}
	\label{fig:3} 
\end{figure}

Out-links of a query term are chosen by extracting the hyperlinks from the Wikipedia page of the query term as shown in Fig. \ref{fig:3}. For example, if the initial query term is ``Bird" then all the hyperlinks within the body of the article ``Bird" are extracted as out-links.

\subsubsection{Assigning in-link score to expansion terms}

After extraction of in-links and out-links of the query term, expansion terms are selected from the out-links on the basis of semantic similarity. The semantic similarity has been calculated based on in-link scores. Let $t$ be a query term and $t_1$ be  one of its candidate expansion terms. In reference to Wikipedia, these two articles $t$ and $t_1$ are considered to be semantically similar if (i) $t_1$ is both an out-link and an in-link of $t$, and (ii) $t_1$ has a high in-link score. The in-link score is based on the popular weighting scheme $tf\:.\:idf$ in IR and is calculated as follows:
\begin{equation}\label{eq:3}
Score(I(t_1))= tf(t,t_1) \:.\: idf(t_1,W_D)
\end{equation} 
where:\\ $tf(t,t_1)$ is the term frequency of `query term $t$ and its synonyms obtained from WordNet' in the article $t_1$, and\\
$idf(t_1,W_D)$ is the inverse document frequency of term $t_1$ in the whole of Wikipedia dump $W_D$.\\
$idf$ can be calculated as follows:
\begin{equation}\label{eq:4}
idf(t_1,W_D)=log \frac{N}{|\{d \in W_D : t_1 \in d\}|}
\end{equation}
where:\\$N$ is the total number of articles in the Wikipedia dump, and\\$|\{d \in W_D : d \in t_1\}|$ is the number of articles where the term $t_1$ appears.

The intuition behind the in-link score is to capture (1) the amount of similarity between the expansion term and the initial query term, and (2) the amount of useful information being provided by the expansion term with respect to QE, i.e, whether the expansion term is common or rare across the whole Wikipedia dump.

Elaborating on the above two points, the term frequency $tf$ provides the semantic similarity between the initial query term and the expansion term, whereas $idf$ provides a score for the rareness of an expansion term. The latter assigns lower priority to the stop words (common terms) in Wikipedia articles (e.g., Main Page, Contents, Edit, References, Help, About Wikipedia, etc.). In Wikipedia, both common terms and expansion terms are hyperlinks of the query term article; the $idf$ helps in removing these common hyperlinks that are present in all the articles of the candidate expansion terms. 

After assigning an in-link score to each expanded term, for each term in the initial query, we select top $n$ terms based on their in-link scores. These top $n$ terms form the intermediate expanded query. After this, these intermediate terms are re-weighted using a correlation score (as described in Sec. \ref{correlation}). The top $m$ terms chosen on the basis of the correlation score become one part of the expanded query. The other part is obtained from WordNet as described next. 
\subsection{QE using WordNet}
After preprocessing of the initial query, the individual terms and phrases obtained as keywords are searched in WordNet for QE. While extracting semantically similar terms from WordNet, more priority is given to the phrases in the query than the individual terms. Specifically, phrases (formed by two consecutive words) are looked up first in WordNet for expansion. Only when no entity is found in WordNet corresponding to a phrase are its individual terms looked up separately in WordNet. It should be noted that phrases are considered only at the time of finding semantically similar terms from WordNet.

\begin{figure}[h]
	\centering 
	\includegraphics[width=13cm, height=7cm]{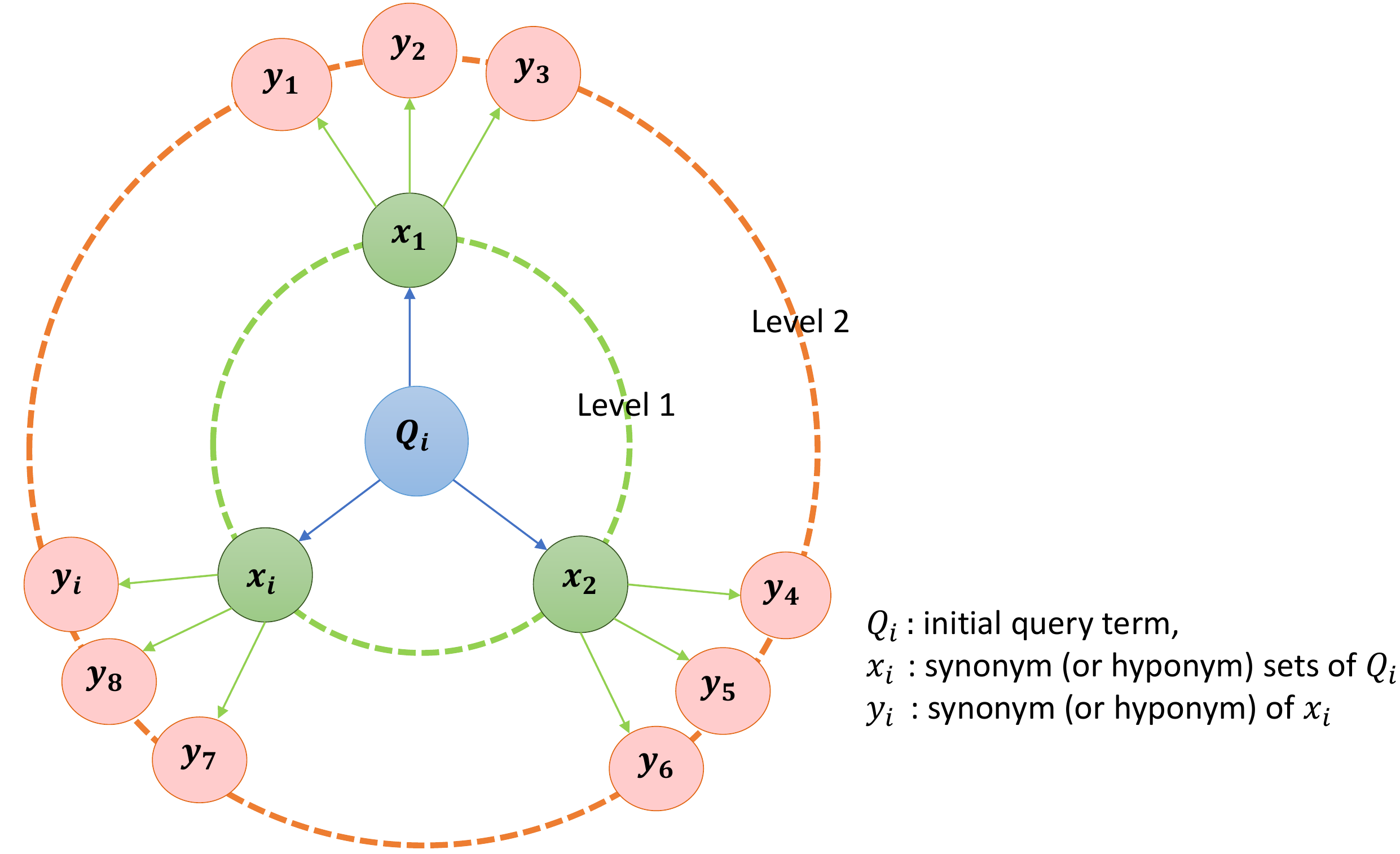}  
	%\captionsetup{justification=centering}    
	\caption{Initial query term and its two-level synonym (or hyponym) sets}
	\label{fig:4} 
\end{figure}

When querying for semantically similar terms from WordNet, only synonym and hyponym sets of the query term are considered as candidate expansion terms. Here synonyms and hyponyms are fetched at two levels, i.e., as shown in Fig.\ref{fig:4} for an initial query term $Q_i$, at level one its synonyms, denoted through $x_i$s, are considered, and, at level two, synonyms of  $x_i$s, denoted through $y_i$s, are considered. The final synonym set used for QE is the union of level one and level two synonyms. Similarly, hyponyms are also fetched at two levels.

After fetching synonyms and hyponyms at two levels, a wide range of semantically similar terms are obtained. Next, we rank these terms using  $tf\:.\:idf$:
\begin{equation}\label{eq:5}
Score(t_1)= tf(t_1,t) \:.\: idf(t_1,W_D)
\end{equation} 
where:\\ $t$ is the initial query term,\\ $t_1$ is an expanded term,\\ $tf(t_1,t)$ is the term frequency of expanded term $t_1$ in the Wikipedia article of query term $t$, and\\ $idf(t_1, W_D)$ is the inverse document frequency of term $t_1$ in the whole Wikipedia dump $W_D$.\\ $idf$ is calculated  as given in Eq. \ref{eq:4}.

After ranking expanded terms based on the above score, we collect the top $n$ terms as the intermediate expanded query. These intermediate terms are re-weighted using the correlation score. Top $m$ terms chosen on the basis of the correlation score (as described in Sec. \ref{correlation}) become the second part of the expanded query, the first part being obtained from Wikipedia as described before. 

\subsection{Re-weighting Expanded Terms} 
\label{correlation}
So far, a set of candidate expansion terms have been obtained, where each expansion term is strongly connected to an individual query term or phrase. These terms have been assigned weights using in-link score (for terms obtained from Wikipedia) and $tf\:.\:idf$ score (for terms obtained from WordNet). However, this may not properly capture the relationship of the expansion term to the query as a whole. For example, the word ``technology" is frequently associated with the word ``information''.  Here, expanding the query term ``technology" with ``information" might work well for some queries such as ``engineering technology", ``science technology", and ``educational technology" but might not work well for others such as ``music technology", ``food technology", and ``financial technology". This problem has also been discussed in reference \cite{bai2007using}. To resolve this language ambiguity problem, we re-weight expanded terms using the correlation score \cite{xu1996query}. The logic behind doing so is that if an expansion feature is correlated to several individual query terms, then the chances are high that it will be correlated to the query as a whole as well.

The correlation score is described as follows. Let $q$ be the original query and let $t_1$ be a candidate expansion term. The correlation score of $t_1$ with $q$ is calculated as: 
\begin{equation}
\begin{split}
C_{t_1, q} &=\frac{1}{|q|}\sum_{t\in q}\, c_{t, t_1} \\
& =\frac{1}{|q|}\sum_{t\in q}\, w_{t, a_t} \:.\: w_{t_1, a_t}
\end{split}
\label{reweight}
\end{equation}
where:\\ $c_{t, t_1}$ denotes the correlation (similarity) score between terms $t$ and $t_1$, and\\ $w_{t, a_t}$ ($w_{t_1, a_t}$) is the weight of the term $t$ ($t_1$) in the article $a_t$ of term $t$.\\
The weight of the term $t$ in its article $t$$(a_t)$, denoted $w_{t, a_t}$ ($w_{t_1, a_t}$ is similarly defined),  is computed as:
\begin{equation}
\begin{split}
w_{t, a_t} &= tf(t, a_t) \:.\: itf(t ,a_q) \\
& =tf(t, a_t) \:.\: log \frac{T}{|T_{a_t}|}
\end{split}
\label{itf}
\end{equation}
where:\\ $ tf(t, a_t)$ is the term frequency of term $t$ in its article $a_t$,
\\ $a_q$ denotes all Wikipedia articles corresponding to the terms in the original query $q$, 
\\ $itf(t, a_q)$ is the inverse term frequency of term $t$ associated with $a_q$,
\\ $T$ is the frequency of term $t$ in all the Wikipedia articles in set  $a_q$, and
\\ $|T_{a_t}|$ is the frequency of term $t$ in the article $a_t$. 

After assigning the correlation score to expansion terms, we collect the top $m$ terms from both data sources  to form the final set of expanded terms.

\section{Experimental Setup}
\label{Experimental Setup}
In order to evaluate the proposed WWQE approach, the experiments were carried out on a large number of queries (50) from FIRE ad-hoc test collections. As real-life queries are short, we used only the title field of all queries. We used  Brill's tagger to assign a POS tag to each query term for extracting the phrase and individual word. These phrases and individual words have been used for QE. We used the most recent Windows version of WordNet 2.1 to extract two levels of synsets terms and Wikipedia for in-links extraction for QE. 

We use the Wikipedia Dump (also known as `WikiDump') for in-link extraction. Wikipedia dump contains every Wikipedia article in XML format. As an open source project, the Wikipedia dump can be downloaded from \emph{https://dumps.wikimedia.org/}. We downloaded the English Wikipedia dump titled ``enwiki-20170101-pages-articles-multistream.xml" of January 2017. 

We compared the performance of our query expansion technique with several existing weighting models and other related state-of-the-art techniques as described in Sec.\ref{Evaluation Metrics}.

\subsection{Dataset}
We use the well-known benchmark dataset Forum for Information Retrieval Evaluation (FIRE\footnote{http://fire.irsi.res.in/fire/static/data}) to evaluate our proposed WWQE approach. Table \ref{tab1} summarizes the dataset used. FIRE collections consists of a very large set of documents on which IR is done, a set of questions (called topics), and the right answers (called relevance judgments) stating the relevance of  documents to the corresponding topic(s). The FIRE dataset consists of a large collection of newswire articles from two sources, namely BDnews24\footnote{https://bdnews24.com/} and The Telegraph\footnote{https://www.telegraphindia.com/}, provided by Indian Statistical Institute Kolkata, India.
\begin{table}[!h]
	%\begin{center}
	\centering
	\caption{Statistics of experimental corpora \label{tab1}}{
		
		\begin{tabular}{ | M{2.2cm} | M{2cm} | M{2cm} | M{2cm} | M{2cm} | }
			\hline

			\textbf{Corpus} & \textbf{ Source} & \textbf{Size} & \textbf{\# of docs} & \textbf{Queries} \\ \hline 
			FIRE  & FIRE 2011 (English) & 1.76 GB & 3,92,577 & 126 - 175 \\ \hline
			
	\end{tabular}}
	%\end{center}
\end{table}	
\subsection{Evaluation Metrics}	
\label{Evaluation Metrics}
We used the TERRIER\footnote{http://terrier.org/} retrieval system for all our experimental evaluations. We used the title field of the topics in the FIRE dataset. For indexing the documents, first the stopwords were removed, and then Porter's Stemmer was used for stemming the documents. All experimental evaluations are based on the unigram word assumption, i.e., all documents and queries in the corpus are indexed using single terms. We did not use any phrase or positional information. To compare the effectiveness of our expansion technique, we used the following weighting models: IFB2, a probabilistic divergence from randomness (DFR) model \cite{amati2002probabilistic}, BM25 model of Okapi \cite{robertson1996okapi}, Laplace's law of succession I(n)L2 \cite{good1965estimation}, Log-logistic DFR model LGD \cite{clinchant2010information}, Divergence from Independence model DPH \cite{amati2008fub},  and Standard tf-idf model. The parameters for these models were set to the default values in TERRIER.

We evaluated the  results on standard evaluation metrics: MAP (mean average precision), GM\_MAP (geometric mean average precision), F-Measure, P@10 (precision at top 10 ranks), P@20, P@30, bpref (binary preference) and the overall recall (number of relevant documents retrieved). Additionally, we reported the percentage improvement in MAP over the baseline (non-expanded query) for each expansion method and other related methods. We also investigated the retrieval effectiveness of the proposed technique with a number of expansion terms.

\section{Experimental Results}
\label{Experimental Results}
The aim of our experiments is to explore the effectiveness of the proposed Wikipedia-WordNet-based QE technique (WWQE) by comparing it with the three baselines on popular weighting models and evaluation metrics. The three baseline approaches are: (i) unexpanded query, (ii) query expansion using Wikipedia alone, and (iii) query expansion using WordNet alone. Comparative analysis is shown in Tables \ref{QE using Wikipedia alone}, \ref{QE using WordNet alone} and \ref{tab2}.
  
\begin{table}[!h]
	%\begin{center}
	\centering
	\caption{Comparison of QE using Wikipedia alone on popular models with the top 30 expansion terms on the FIRE dataset \label{QE using Wikipedia alone}}{
		\begin{tabular}{ |p{1.4cm}||p{1.5cm}|p{1.5cm}|p{1.5cm}|p{1.5cm}|p{1.5cm}|p{1.5cm}|  }
			\hline
			\multicolumn{7}{|c|}{\textbf{Model Performance without Query Expansion}} \\
			\hline
			Method & MAP & GM\_MAP & P@10 & P@20 & P@30 & \#rel\_ret \\
			\hline
			IFB2  & 0.2765 & 0.1907 & 0.3660 & 0.3560 & 0.3420 & 2330\\
			I(n)L2& 0.2979 & 0.2023 & 0.4280 & 0.3900 & 0.3553 &  2322\\
			LGD & 0.2909 & 0.1974 & 0.4100 & 0.3710 & 0.3420 & 2309\\
			DPH & 0.3133 & 0.2219 & 0.4540 & 0.4040 & 0.3653 & 2338\\
			BM25 & 0.3163 & 0.2234 & 0.4600 & 0.3970 & 0.3660 &  2343\\
			Tf-idf & 0.3183 & 0.2261 & 0.4560 & 0.4010 & 0.3707 & 2340\\ \hline \hline
			\multicolumn{7}{| c |}{\textbf{Model Performance with QE using Wikipedia alone}}\\ \hline
			IFB2  & 0.3166 \textbf{($\uparrow$14.5\%)}& 0.2498 \textbf{($\uparrow$30.99\%)} & 0.4162 \textbf{($\uparrow$13.72\%)} & 0.3969 \textbf{($\uparrow$11.49\%)} & 0.3623 ($\uparrow$5.94\%) & 2420 ($\uparrow$3.86\%)\\
			I(n)L2& 0.3317 ($\uparrow$11.35\%) & 0.2628 ($\uparrow$29.91\%) &  0.4425 ($\uparrow$3.39\%) & 0.4012 ($\uparrow$2.87\%) & 0.3892 \textbf{($\uparrow$9.54\%)}&  2432 ($\uparrow$4.74\%)\\
			LGD & 0.3248 ($\uparrow$11.65\%) & 0.2535 ($\uparrow$28.42\%) & 0.4432 ($\uparrow$8.1\%) & 0.3901 ($\uparrow$5.15\%) & 0.3639 ($\uparrow$6.4\%)& 2428 \textbf{($\uparrow$5.15\%)}\\
			DPH & 0.3291 ($\uparrow$5.04\%) & 0.2598 ($\uparrow$17.08\%) & 0.4667 ($\uparrow$2.8\%) & 0.4127 ($\uparrow$2.15\%) & 0.3783 ($\uparrow$3.56\%)& 2423 ($\uparrow$3.64\%)\\
			BM25 & 0.3304 ($\uparrow$4.46\%) & 0.2501 ($\uparrow$11.95\%) &  0.4723 ($\uparrow$2.67\%) & 0.4044 ($\uparrow$1.86\%) & 0.3717 ($\uparrow$1.56\%) &  2421 ($\uparrow$3.33\%)\\
			Tf-idf & 0.3315 ($\uparrow$4.15\%) & 0.2572 ($\uparrow$13.75\%) & 0.4691 ($\uparrow$2.87\%) & 0.4123 ($\uparrow$2.82\%) & 0.3875 ($\uparrow$4.53\%)& 2422 ($\uparrow$3.5\%)\\
			\hline
	\end{tabular}}
\end{table}

\begin{table}[!h]
	%\begin{center}
	\centering
	\caption{Comparison of QE using WordNet alone on popular models with the top 30 expansion terms on the FIRE dataset \label{QE using WordNet alone}}{
		\begin{tabular}{ |p{1.4cm}||p{1.5cm}|p{1.5cm}|p{1.5cm}|p{1.5cm}|p{1.5cm}|p{1.5cm}|  }
			\hline
			\multicolumn{7}{|c|}{\textbf{Model Performance without Query Expansion}} \\
			\hline
			Method & MAP & GM\_MAP & P@10 & P@20 & P@30 & \#rel\_ret \\
			\hline
			IFB2  & 0.2765 & 0.1907 & 0.3660 & 0.3560 & 0.3420 & 2330\\
			I(n)L2& 0.2979 & 0.2023 & 0.4280 & 0.3900 & 0.3553 &  2322\\
			LGD & 0.2909 & 0.1974 & 0.4100 & 0.3710 & 0.3420 & 2309\\
			DPH & 0.3133 & 0.2219 & 0.4540 & 0.4040 & 0.3653 & 2338\\
			BM25 & 0.3163 & 0.2234 & 0.4600 & 0.3970 & 0.3660 &  2343\\
			Tf-idf & 0.3183 & 0.2261 & 0.4560 & 0.4010 & 0.3707 & 2340\\ \hline \hline
			\multicolumn{7}{| c |}{\textbf{Model Performance with QE using WordNet alone}}\\ \hline
			IFB2  & 0.2901 ($\uparrow$4.92\%) & 0.2113 ($\uparrow$10.8\%) & 0.3817 \textbf{($\uparrow$4.29\%)} & 0.3693 ($\uparrow$3.74\%) & 0.3521 ($\uparrow$2.95\%) & 2361 ($\uparrow$1.33\%)\\
			I(n)L2& 0.3112 ($\uparrow$4.46\%) & 0.2246 \textbf{($\uparrow$11.02\%)} &  0.4373 ($\uparrow$2.17\%) & 0.3972 ($\uparrow$1.85\%) & 0.3648 ($\uparrow$2.67\%) &  2358 \textbf{($\uparrow$1.55\%)}\\
			LGD & 0.3101 \textbf{($\uparrow$6.6\%)} & 0.2177 ($\uparrow$10.28\%) & 0.4111 ($\uparrow$0.27\%) & 0.3872 \textbf{($\uparrow$4.37\%)} & 0.3513 ($\uparrow$2.72\%) & 2327 ($\uparrow$0.78\%)\\
			DPH & 0.3178 ($\uparrow$1.43\%) & 0.2295 ($\uparrow$3.42\%) & 0.4627 ($\uparrow$1.92\%) & 0.4105 ($\uparrow$1.61\%) & 0.3712 ($\uparrow$1.62\%)& 2359 ($\uparrow$0.89\%)\\
			BM25 & 0.3199 ($\uparrow$1.14\%) & 0.2301 ($\uparrow$3\%) &  0.4612 ($\uparrow$0.26\%) & 0.3999 ($\uparrow$0.73\%) & 0.3725 ($\uparrow$1.78\%) &  2353 ($\uparrow$0.43\%)\\
			Tf-idf & 0.3203 ($\uparrow$0.63\%) & 0.2312 ($\uparrow$2.26\%) & 0.4597 ($\uparrow$0.84\%) & 0.4098 ($\uparrow$2.19\%) & 0.3827 \textbf{($\uparrow$3.24\%)} & 2345 ($\uparrow$0.21\%)\\
			\hline
	\end{tabular}}
\end{table}

\begin{table}[!h]
	%\begin{center}
	\centering
	\caption{Comparison of the proposed WWQE technique on popular models with the top 30 expansion terms on the FIRE dataset \label{tab2}}{
		\begin{tabular}{ |p{1.4cm}||p{1.5cm}|p{1.5cm}|p{1.5cm}|p{1.5cm}|p{1.5cm}|p{1.5cm}|  }
			\hline
			\multicolumn{7}{|c|}{\textbf{Model Performance without Query Expansion}} \\
			\hline
			Method & MAP & GM\_MAP & P@10 & P@20 & P@30 & \#rel\_ret \\
			\hline
			IFB2  & 0.2765 & 0.1907 & 0.3660 & 0.3560 & 0.3420 & 2330\\
			I(n)L2& 0.2979 & 0.2023 & 0.4280 & 0.3900 & 0.3553 &  2322\\
			LGD & 0.2909 & 0.1974 & 0.4100 & 0.3710 & 0.3420 & 2309\\
			DPH & 0.3133 & 0.2219 & 0.4540 & 0.4040 & 0.3653 & 2338\\
			BM25 & 0.3163 & 0.2234 & 0.4600 & 0.3970 & 0.3660 &  2343\\
			Tf-idf & 0.3183 & 0.2261 & 0.4560 & 0.4010 & 0.3707 & 2340\\ \hline \hline
			\multicolumn{7}{| c |}{\textbf{Model Performance with Proposed Query Expansion Technique}}\\ \hline
			IFB2  & 0.3439 \textbf{($\uparrow$24.38\%)}& 0.2835 \textbf{($\uparrow$48.66\%)} & 0.4660 \textbf{($\uparrow$27.49\%)} & 0.4400 \textbf{($\uparrow$23.60\%)} & 0.4040 ($\uparrow$18.13\%) & 2554 ($\uparrow$9.61\%)\\
			I(n)L2& 0.3552 ($\uparrow$19.23\%) & 0.2933 ($\uparrow$44.98\%) &  0.4900 ($\uparrow$14.48\%) & 0.4560 ($\uparrow$16.92\%) & 0.4200 ($\uparrow$18.21\%)&  2583 \textbf{($\uparrow$11.24\%)}\\
			LGD & 0.3460 ($\uparrow$18.94\%) & 0.2855 ($\uparrow$44.63\%) & 0.4900 ($\uparrow$19.51\%) & 0.4460 ($\uparrow$20.21\%) & 0.4187 \textbf{($\uparrow$22.43\%)}& 2566 ($\uparrow$11.13\%)\\
			DPH & 0.3497 ($\uparrow$11.62\%) & 0.2902 ($\uparrow$30.78\%) & 0.4940 ($\uparrow$8.81\%) & 0.4490 ($\uparrow$11.14\%) & 0.4113 ($\uparrow$12.59\%)& 2565 ($\uparrow$9.71\%)\\
			BM25 & 0.3508 ($\uparrow$10.91\%) & 0.2878 ($\uparrow$28.83\%) &  0.5160 ($\uparrow$12.17\%) & 0.4490 ($\uparrow$13.10\%) & 0.4093 ($\uparrow$11.83\%) &  2560 ($\uparrow$9.26\%)\\
			Tf-idf & 0.3521 ($\uparrow$10.62\%) & 0.2896 ($\uparrow$27.95\%) & 0.5100 ($\uparrow$11.84\%) & 0.4520 ($\uparrow$12.72\%) & 0.4120 ($\uparrow$11.14\%)& 2561 ($\uparrow$9.44\%)\\
			\hline
	\end{tabular}}
\end{table}

Table \ref{tab2} shows a performance comparison of the proposed WWQE technique over popular weighting models in the context of MAP, GM\_MAP, P@10, P@20, P@30, and relevant return. The table shows that the proposed WWQE technique  is compatible with the existing popular weighting models and it also improves the information retrieval effectively. It also shows the relative percentage improvements (within parentheses) of various standard evaluation metrics measured against no expansion. By using the proposed query expansion technique (WWQE), the weighting models improve the MAP up to 24\% and GM\_MAP by 48\%. Based on the results presented in Table \ref{tab2} we can say that in the context of all evaluation parameters, the proposed QE technique performs well with all weighting models.\\

\begin{figure}[!h]%
 	\centering
 	\subfloat[]{{\includegraphics[width=7.5cm]{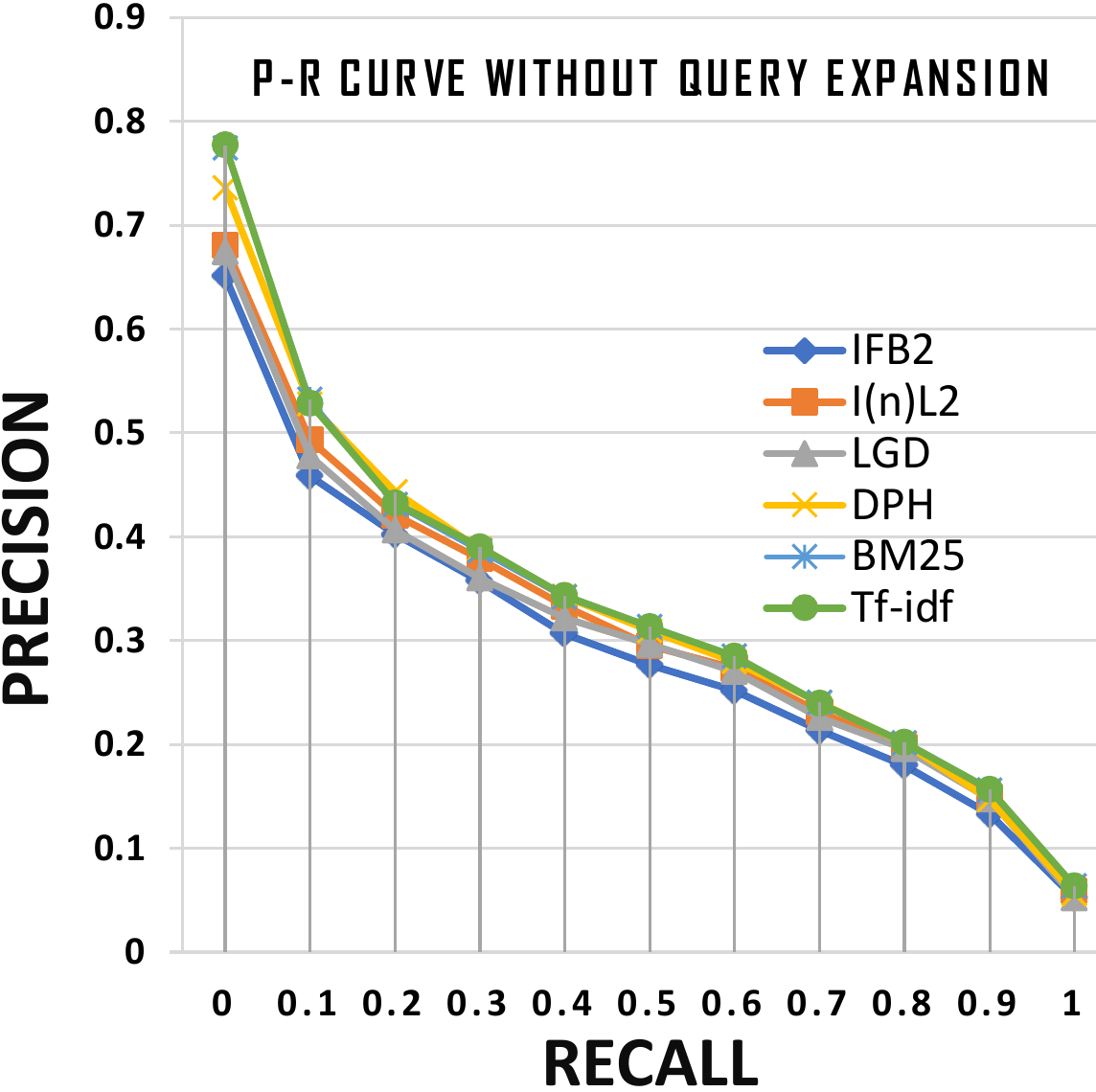} \label{fig:p-r curve without QE} }}%
 	%\label{fig:p-r curve without QE}
 	%\qquad
 	\subfloat[]{{\includegraphics[width=7.5cm]{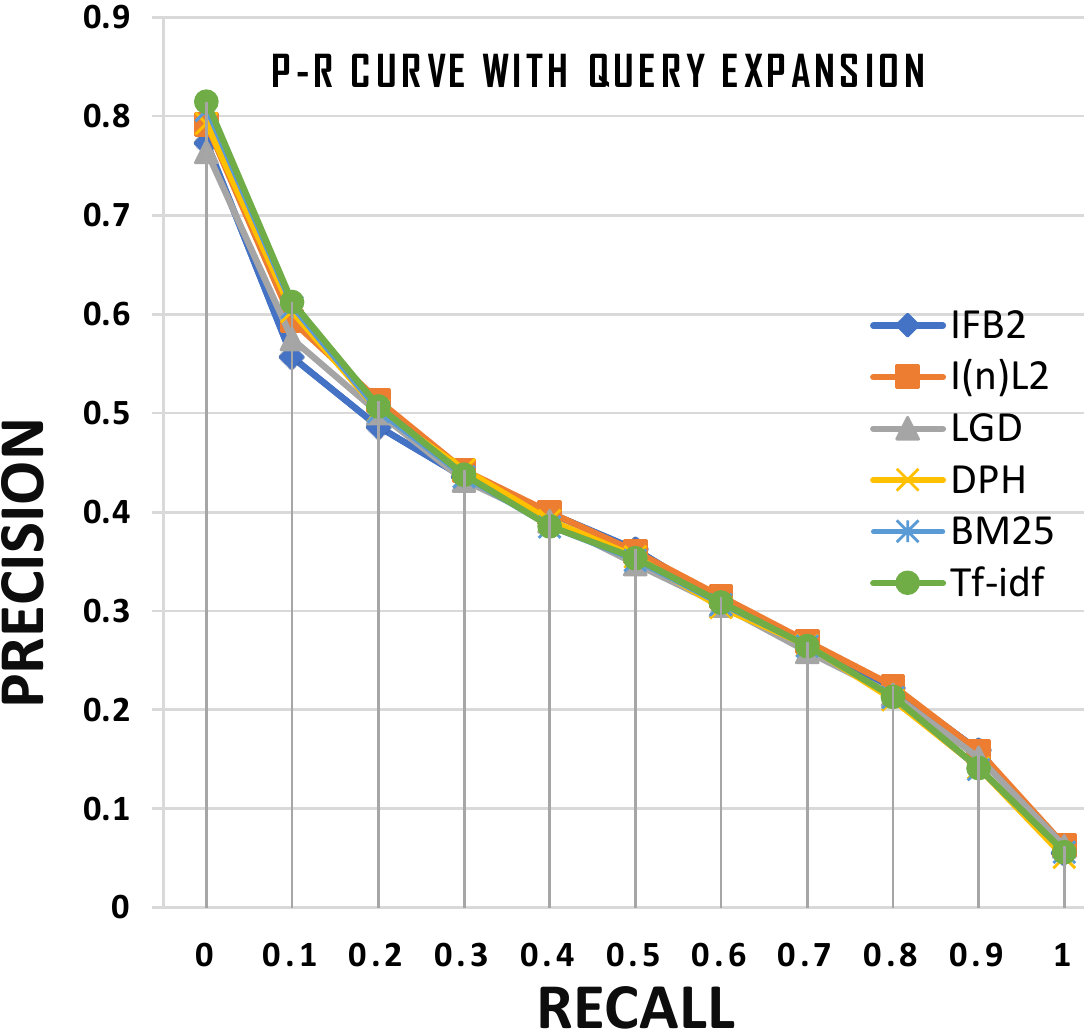} \label{fig:p-r curve with QE} }}%
 	
 	\caption{Comparative analysis of precision-recall curve of proposed QE technique with various weighting models on the FIRE dataset.}%
 	\label{fig:Comparative analysis of Precision-Recall curve}%
 \end{figure}

Figure \ref{fig:Comparative analysis of Precision-Recall curve} shows the comparative analysis of precision-recall curve of WWQE technique with various weighting models. This graph plots the interpolated precision of an IR system using 11 standard cutoff values from the recall levels, i.e., \{0, 0.1, 0.2, 0.3, ...,1.0\}. These graphs are widely used to evaluate IR systems that return ranked documents (i.e., averaging and plotting retrieval results). Comparisons are best made in three different recall ranges: 0 to 0.2, 0.2 to 0.8, and 0.8 to 1. These ranges characterize high precision, middle recall, and high recall performance, respectively.

Based on the graph presented in Figures \ref{fig:p-r curve without QE} and \ref{fig:p-r curve with QE}, we arrive at the conclusion that the P-R curve of the various weighting models have nearly the same retrieval result with or without QE respectively. Therefore we can say that  for improving the information retrieval in QE, the choice of the weighting models is not so important. The importance lies in the choice of technique used for selecting the relevant expansion terms. The relevant expansion terms, in turn,  come from data sources. Hence, the data sources also play an important role for effective QE. This conclusion also supports our proposed WWQE technique, where we select the expansion terms on the basis of individual term weighting as well as assigning a correlation score on the basis of the entire query.

\begin{figure}[h]%
	\centering
	\subfloat[]{{\includegraphics[width=4.8cm]{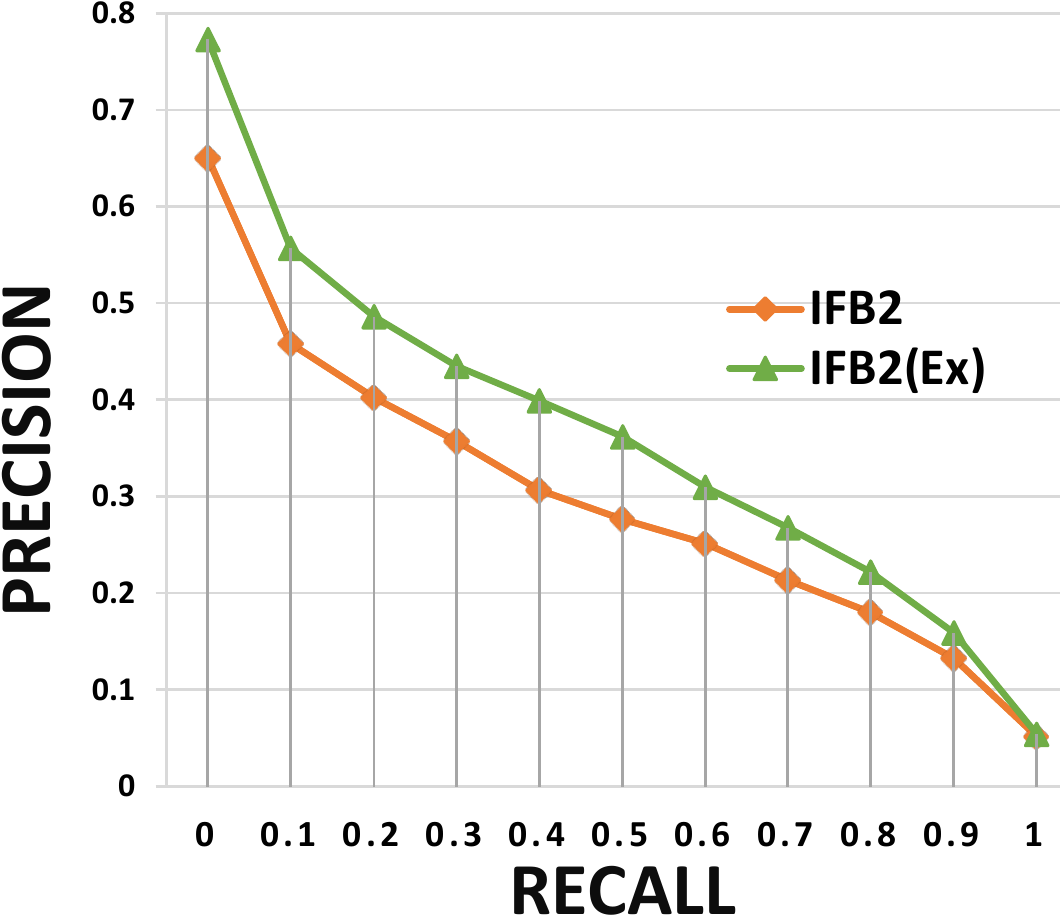} \label{fig:p-r curve IFB2} }}%
	%\qquad
	\subfloat[]{{\includegraphics[width=4.8cm]{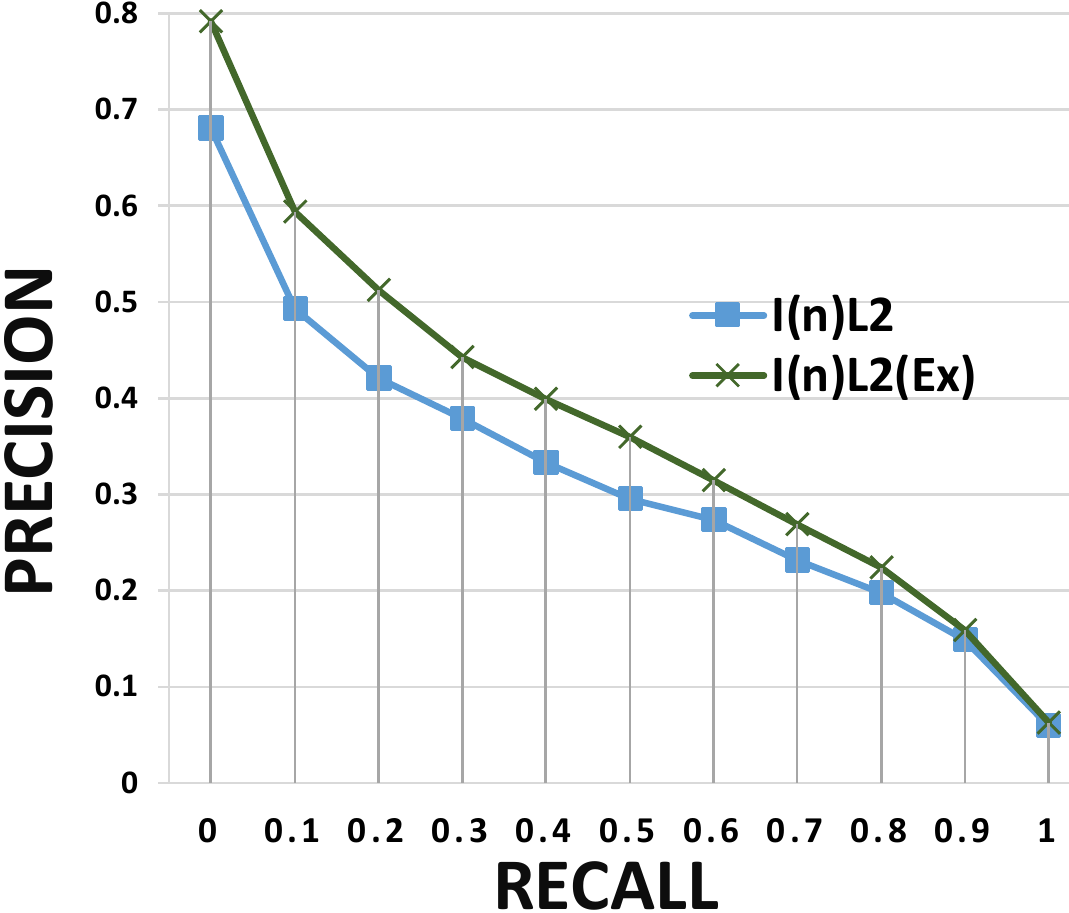} \label{fig:p-r curve InL2} }}%
	%\qquad
	\subfloat[]{{\includegraphics[width=4.8cm]{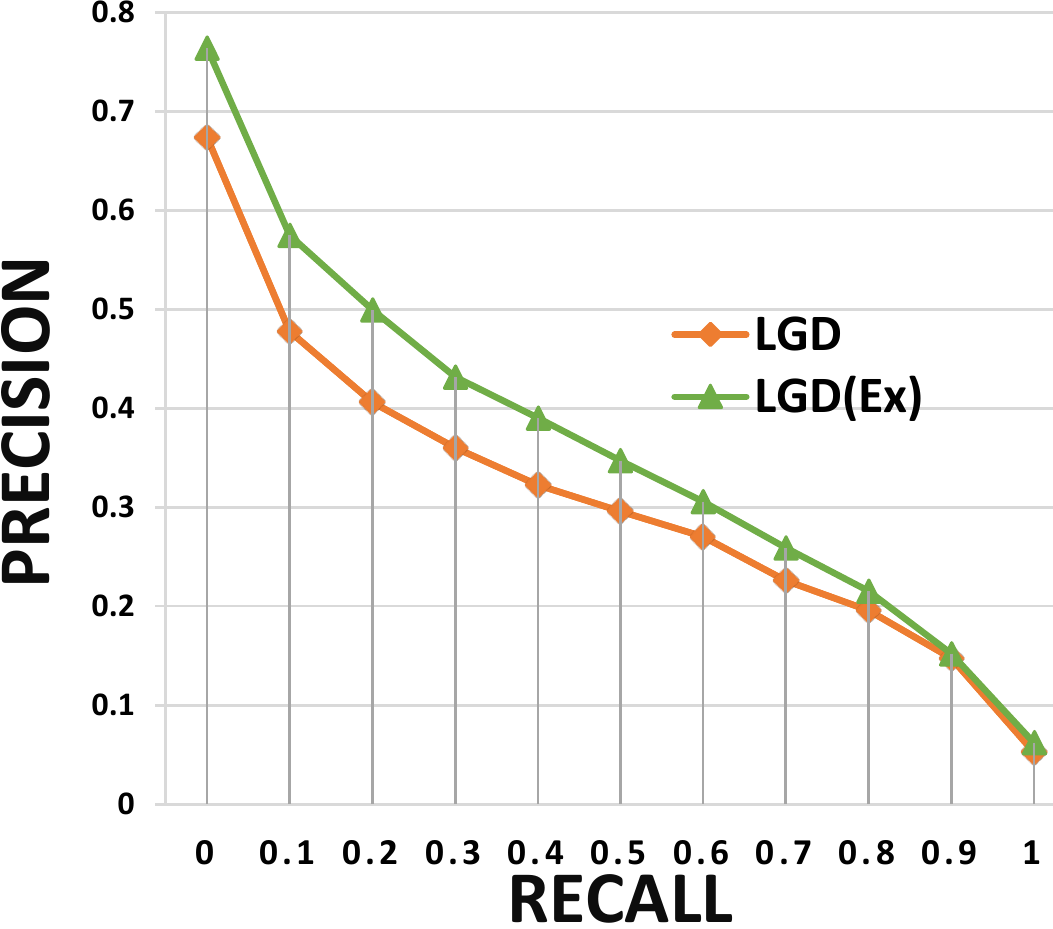}  \label{fig:p-r curve LGD} }}
	\qquad
	\subfloat[]{{\includegraphics[width=4.8cm]{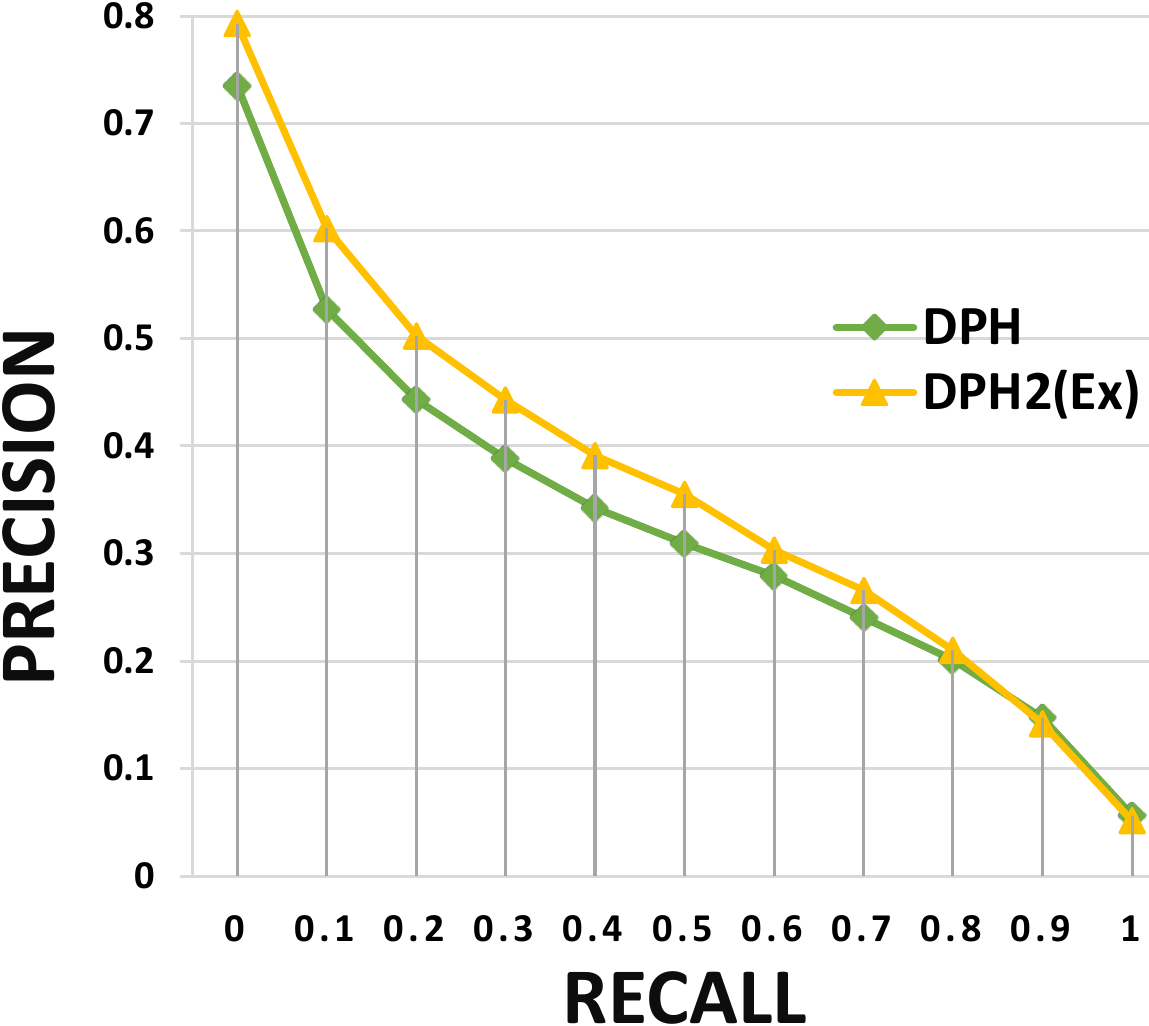}  \label{fig:p-r curve DPH} }}%
	%\qquad
	\subfloat[]{{\includegraphics[width=4.8cm]{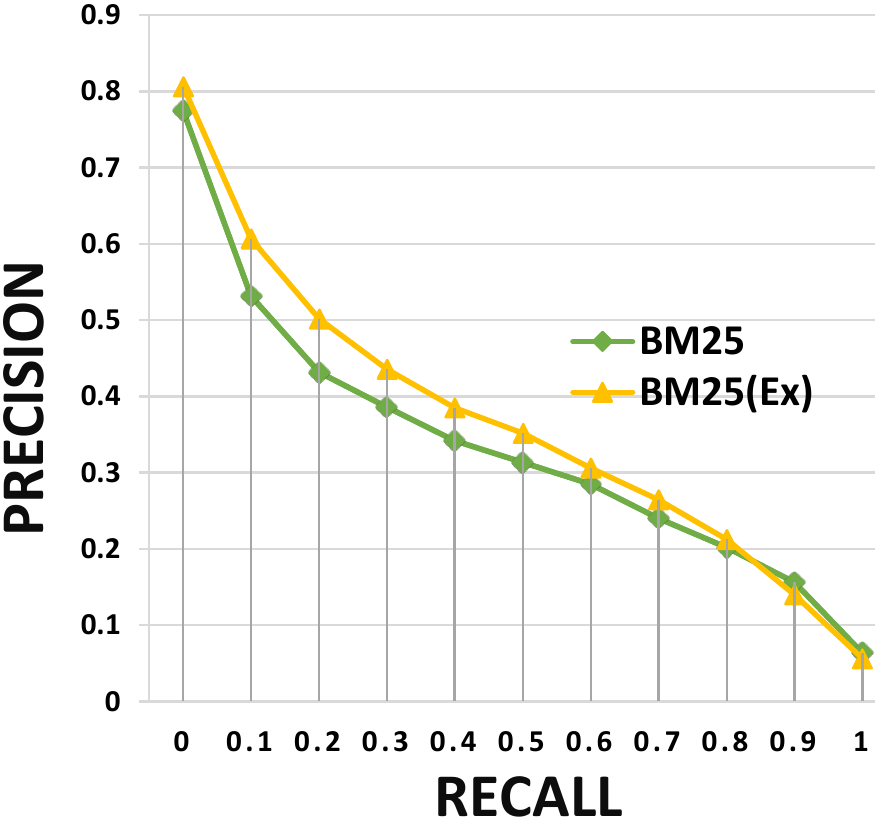} \label{fig:p-r curve BM25} }}%
	%\qquad
	\subfloat[]{{\includegraphics[width=4.8cm]{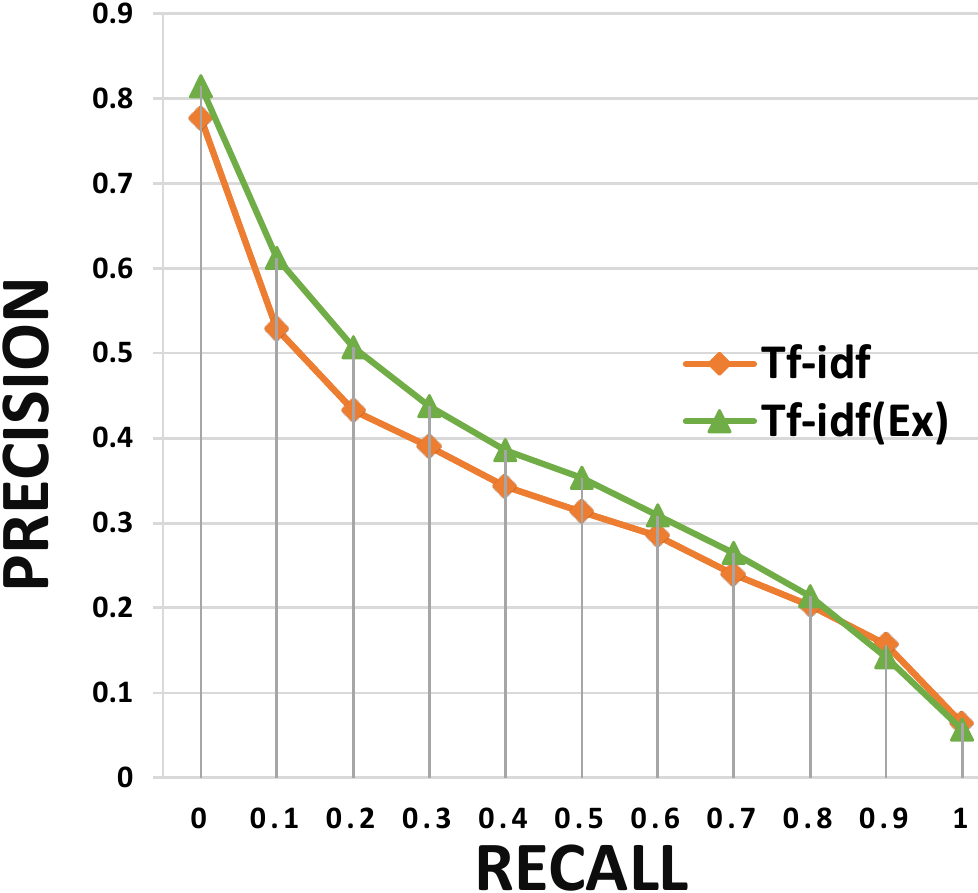} \label{fig:p-r curve TFIDF} }}
	\caption{Comparative analysis of precision-recall curve of WWQE technique with popular weighting models individually on the FIRE dataset. In legend, (Ex) denotes the performance with query expansion, while no parenthesis denotes unexpanded query}%
	\label{fig:Comparative analysis of Precision-Recall curve individually}%
\end{figure}

Figure \ref{fig:Comparative analysis of Precision-Recall curve individually} compares the performance of the WWQE technique with P-R curve using popular weighting models individually. Graphs in the figure show the improvement in retrieval results of the WWQE technique when compared with the initial unexpanded queries.  As indicated in the graph legend, (Ex) denotes the performance with query expansion, while no parenthesis denotes unexpanded query. The P-R curves show the effectiveness of the proposed WWQE technique with all the  popular weighting models. Of all the weighting models, the IFB2 weighting model provides the best retrieval performance with the proposed WWQE technique. 

Figure \ref{fig:Comparative analysis of precision, bpref} compares the WWQE technique in terms of precision, bpref and P@5 with various weighting models on the FIRE dataset in comparison to the unexpanded queries. Here, precision shows the ability of a system to present only relevant documents. P@5 measures the precision over the top 5 documents retrieved and bpref measures a preference relation about how many documents judged as relevant are ranked before the documents judged as irrelevant.

\begin{figure}[!h]%
	\centering
	\subfloat[]{{\includegraphics[width=7.5cm]{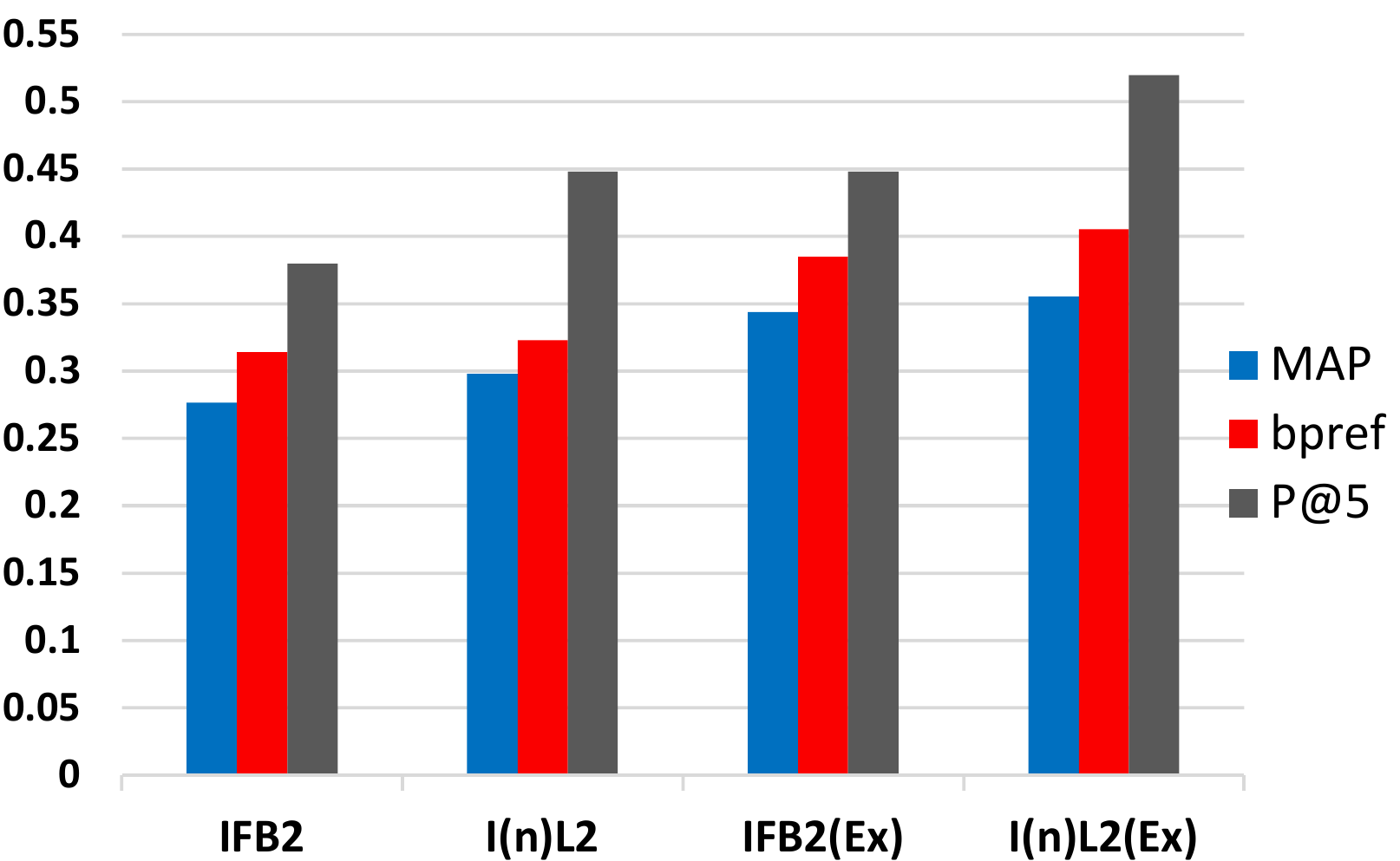}    \label{fig:Chart_IFB2_InL2} }}%
	%\qquad
	\subfloat[]{{\includegraphics[width=7.5cm]{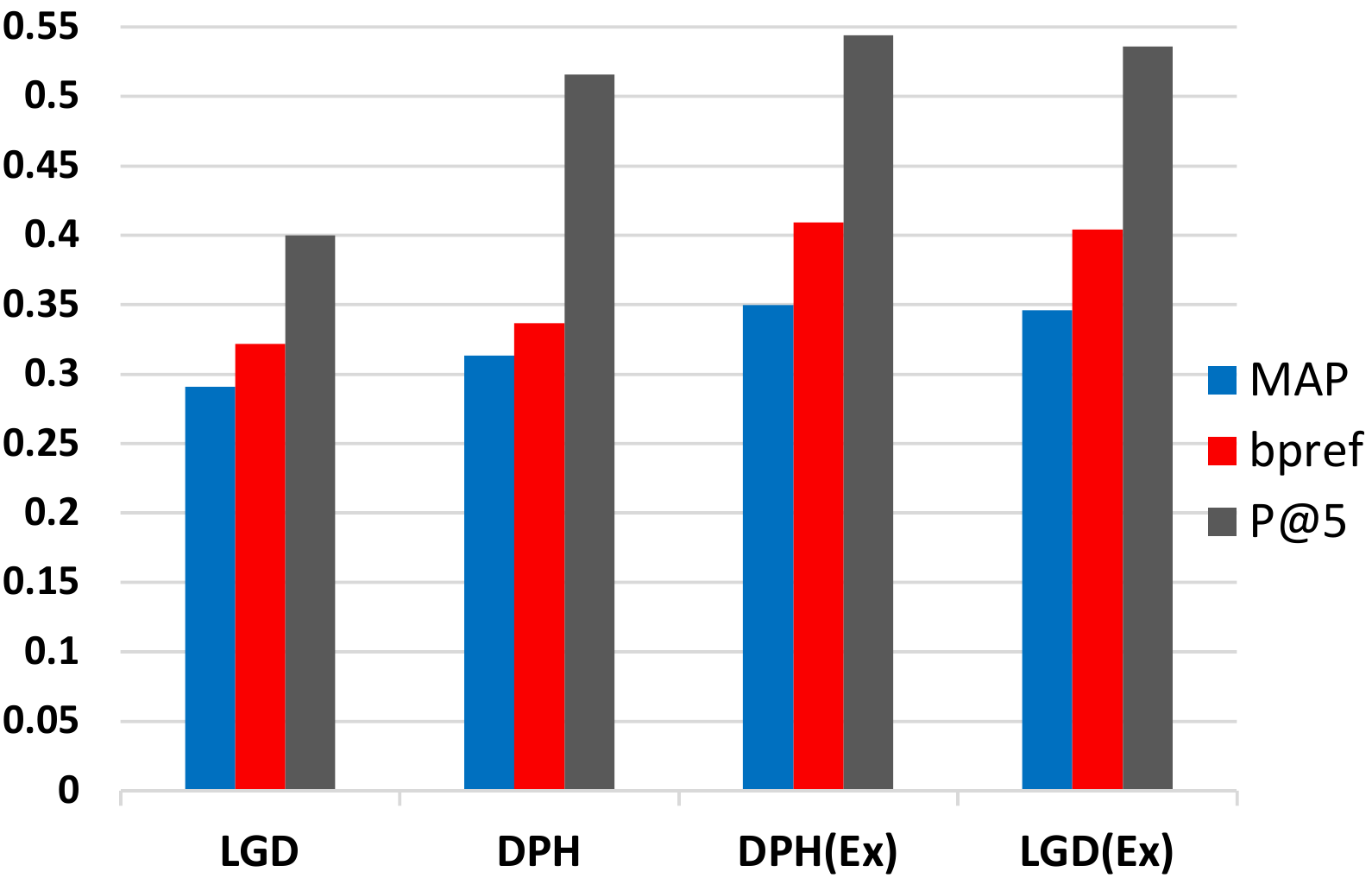}      \label{fig:Chart_LGD_DPH} }}%
	\qquad
	
	\subfloat[]{{\includegraphics[width=7.5cm]{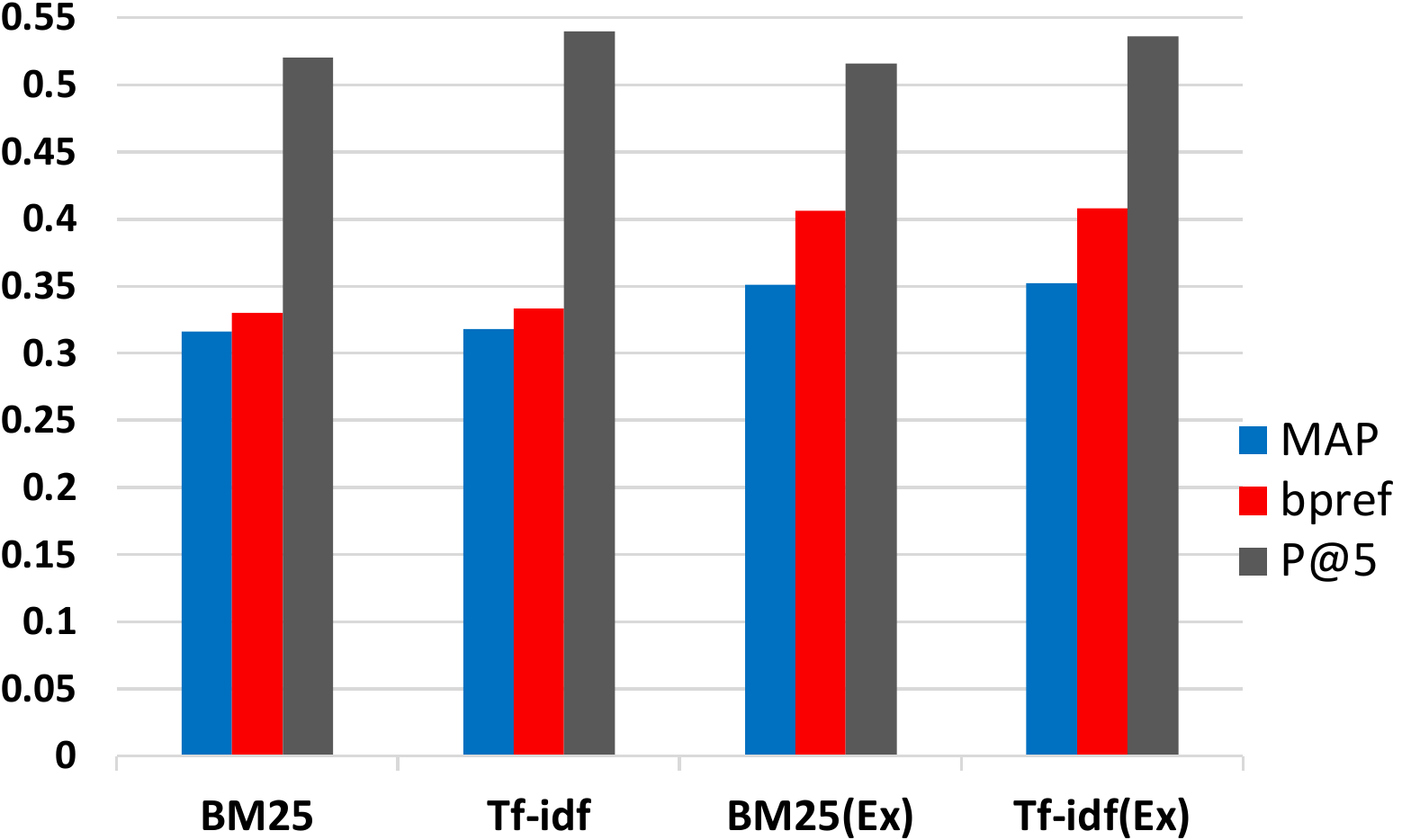}    \label{fig:Chart_BM25_TFIDF} }}
	\caption{Comparative analysis of WWQE technique in terms of precision, bpref and P@5 with various weighting models on the FIRE dataset.}%
	\label{fig:Comparative analysis of precision, bpref}%
\end{figure} 

\begin{figure}[!h]
	\centering 
	\includegraphics[width=11cm, height=7cm]{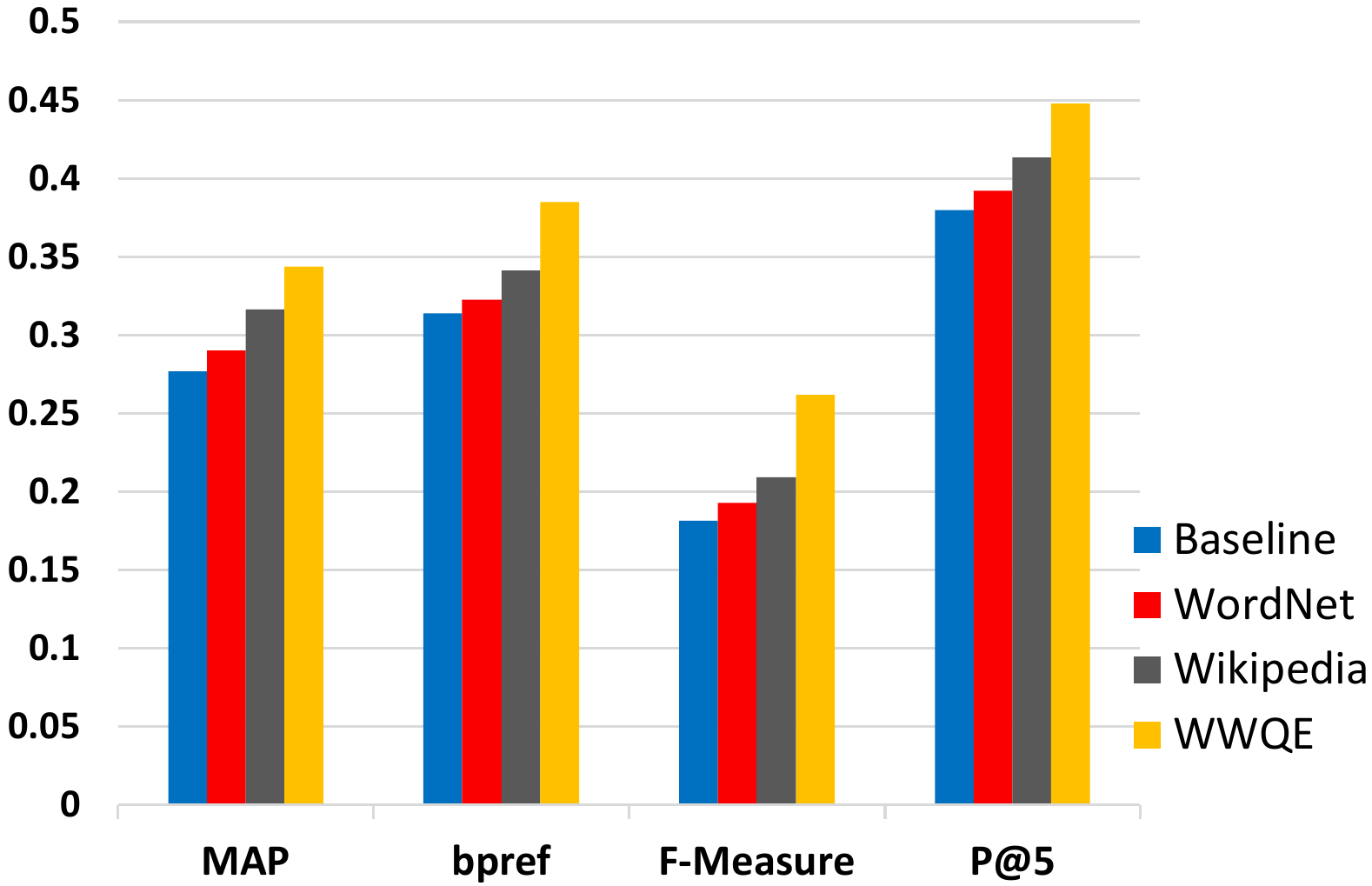}  
	%\captionsetup{justification=centering}    
	\caption{Comparative analysis of WWQE technique with baseline, WordNet and Wikipedia}
	\label{WWQE_W_W} 
\end{figure}

Figure \ref{WWQE_W_W} compares the WWQE technique in terms of MAP, bpref, F-Measure and P@5 with baseline (unexpanded), QE using WordNet alone, and QE using Wikipedia alone. IFB2 model is used for term weighting for experimental evaluation. It can also be observed that the retrieval effectiveness of the proposed WWQE technique is superior to the QE using WordNet and Wikipedia alone.

After evaluating the performance of the proposed QE technique on several popular evaluation metrics, it can be concluded that the proposed QE technique (WWQE) performs well with all weighting models on several evaluation parameters. Therefore, the proposed WWQE technique is effective in  improving information retrieval results.

Table \ref{Comparision} presents a comparison of the proposed WWQE technique with Parapar et al. \cite{parapar2014score} and, Singh and Saran \cite{singh2016relevance} models in terms of Mean Average Precision with the top 30 expansion terms on the FIRE dataset. We can observe that the retrieval effectiveness of the proposed WWQE technique is better than Parapar et al.'s and Singh and Saran's model. However, Parapar et al.'s and Singh and Saran's models perform well in comparison to the QE using WordNet and Wikipedia alone.  

Parapar et al. \cite{parapar2014score} present an approach to minimize the number of non-relevant documents in the pseudo-relevant set to deal with the problem of unwanted documents in the pseudo-relevant collection. These unwanted documents adversely affect the selection of expansion terms. To automatically determine the number of documents to be selected for the pseudo-relevance set for each query, they studied the score distributions in the initial retrieval (i.e., documents retrieved in response to the initial query). The goal of their study was to come up with a threshold score to differentiate between relevant and non-relevant documents.  Singh and Saran's \cite{singh2016relevance} method uses a combination of pseudo-relevance feedback-based QE and multiple-term selection methods. They apply the Borda count rank-aggregating method to weigh expansion terms. Finally, the Word2vec approach is used to select semantically similar terms. The evaluation of their approach shows a significant improvement over individual term
selection method.
	
\begin{table}[!h]
	%\begin{center}
	\centering
	\caption{Comparative analysis of the proposed WWQE technique with Parapar et al.'s, and Singh \& Saran's models with respect to the MAP values \label{Comparision}}{
		
		\begin{tabular}{ | M{1.6cm} | M{4.5cm} | M{3.3cm} |  }
			\hline 
			
			\textbf{Data Set} & \textbf{Methods} & \textbf{MAP}  \\ \hline 
			\multirow{6}{2cm}{\centering {FIRE}} & Baseline (IFB2) & 0.2765
			\\\cline{2-3}  & WordNet model alone & 0.2901 (4.92\%)
			\\\cline{2-3}  & Wikipedia model alone & 0.3166 (14.5\%)
			   \\\cline{2-3}  & Parapar et al. model \cite{parapar2014score} & 0.3178 (14.74\%)
			  \\\cline{2-3}  & Singh and Saran model \cite{singh2016relevance}  & 0.3217 (16.34\%)
			 \\\cline{2-3}    & WWQE (Proposed)  & \textbf{0.3439 (24.38\%)} \\ \hline

	\end{tabular}}
	%\end{center}
\end{table}

\begin{figure}[!h]
	\centering 
	\includegraphics[width=12cm, height=7cm]{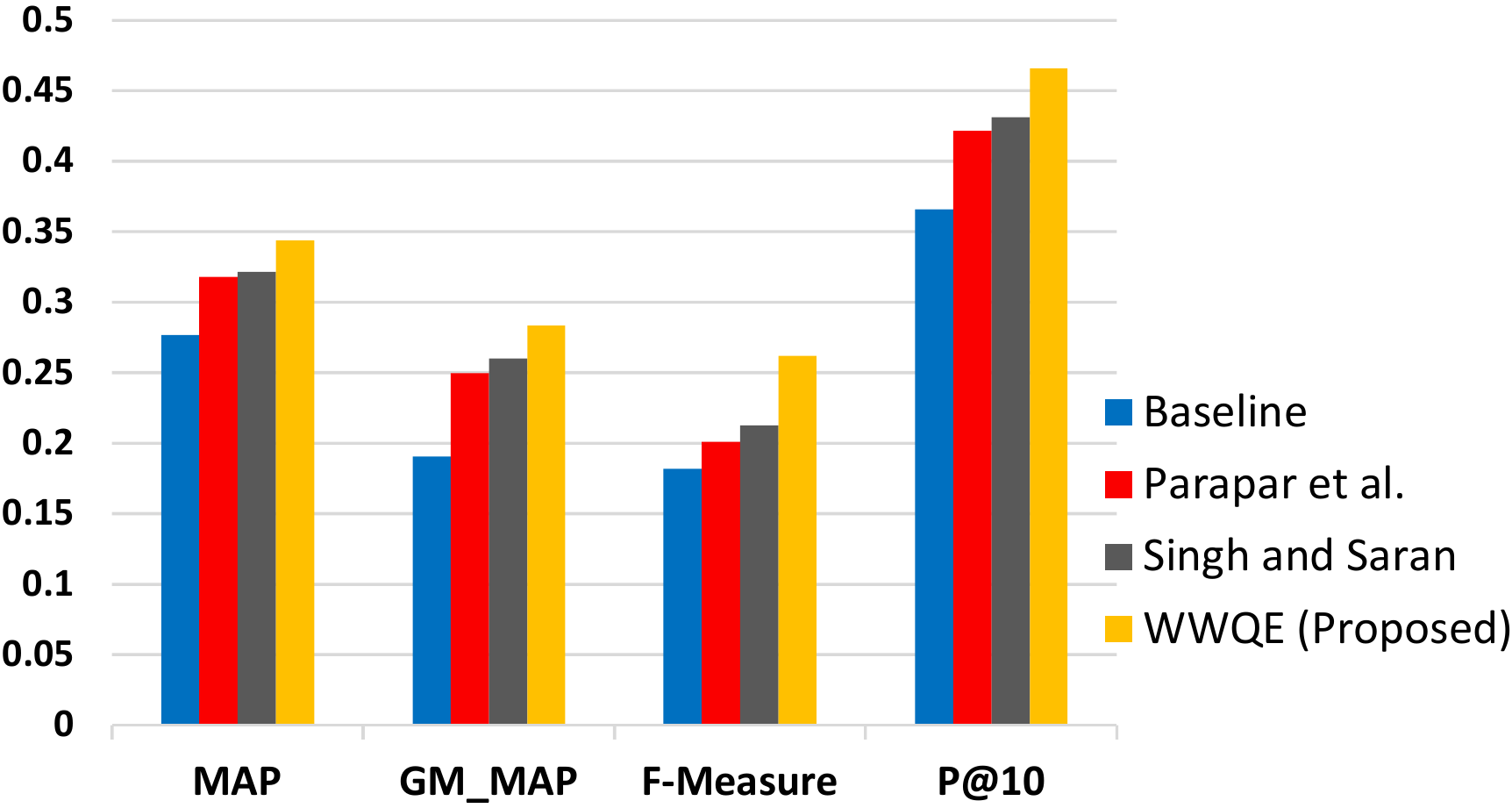}  
	%\captionsetup{justification=centering}    
	\caption{Comparative analysis of the WWQE technique with baseline, WordNet and Wikipedia}
	\label{WWQE_all} 
\end{figure}

Figure \ref{WWQE_all} compares the WWQE technique in terms of the MAP, GM\_MAP, F-Measure and P@10 with baseline (IFB2), Parapar et al.'s, and Singh and Saran's model. It can be clearly seen that the retrieval effectiveness of the proposed WWQE technique is better than the QE using Parapar et al.'s and Singh and Saran's technique.

\subsection{Performance Variation with Number of Expansion Terms}
There are different points of view on the number of expansion terms to be chosen; the number of expansion terms can vary from one-third of the expansion terms to all terms \cite{azad2017query}. However, it might not be realistic to use all of the expansion terms, a small set of expansion terms is usually better than a large set of expansion terms due to noise reduction \cite{salton1990improving}. A limited number of expansion terms may also be important to reduce the response time, especially for a large corpus. However, several studies observed that the number of expansion terms is of low relevance and it varies from query to query \cite{cao2008context}. 

In our experimental results, we show the model performance with the top 30 expansion terms. The number 30 was chosen because it produced better retrieval performance compared to other integral values. We also did experiments by varying the expansion terms from 10 to 50 on our proposed model; Table \ref{Effect of expansion terms} displays the corresponding results. The results show that the variation in performance is very limited with all weighting methods and for all selected sets of expansion terms. However, the retrieval performance of the system is adversely affected when we consider fewer than 20 expansion terms. As mentioned before, for our proposed model, 30 expansion terms seems to give the best results. 

\begin{table}[!h]
	%\begin{center}
	\centering
	\caption{ Effect of the number of expansion terms on the performance (MAP) of the proposed WWQE technique on the FIRE dataset \label{Effect of expansion terms}}{
		\begin{tabular}{ |p{1.4cm}||p{1.5cm}|p{1.4cm}|p{1.4cm}|p{1.4cm}|p{1.4cm}|p{1.4cm}|  }
			\hline
			\multicolumn{7}{|c|}{\textbf{Model Performance vs. Expansion terms}} \\
			\hline
			Method & 10 & 20 & 30 & 40 & 50 & 60 \\
			\hline
			IFB2  & 0.2795 & 0.2907 & \textbf{0.3439} & 0.3401 & 0.3429 & 0.2781\\
			I(n)L2& 0.2997 & 0.2863 & \textbf{0.3552} & 0.3498 & 0.3426 & 0.3345 \\
			LGD  & 0.2981 & 0.3101 & \textbf{0.3460} & 0.3402 & 0.3421 & 0.3234\\
			DPH  & 0.3301 & \textbf{0.3499} & 0.3497 & 0.3498 & 0.3365 & 0.3421\\
			BM25 & 0.3251 & 0.3291 & \textbf{0.3508} & 0.3477 & 0.3379 &  0.3501\\
			Tf-idf & 0.3247 & 0.3365 & 0.3521 & \textbf{0.3610} & 0.3511 & 0.3367\\ \hline 
			
	\end{tabular}}
\end{table}

Table \ref{expansion terms} shows some examples of initial query and its expansion terms obtained with the proposed approached.
\begin{table}[!h]
	%\begin{center}
	\centering
	\caption{Expansion terms obtained with proposed approaches for selected queries on the FIRE dataset \label{expansion terms}}{
		
		\begin{tabular}{ | M{1.3cm} | M{2cm} | M{3.1cm} | M{3.4cm} |M{3.4cm} | }
			\hline

			\textbf{Query ID} & \textbf{Original query} & \textbf{Expansion terms obtained with WordNet} & \textbf{Expansion terms obtained with Wikipedia}  & \textbf{Expansion terms obtained with WWQE} \\ \hline 
			126 & Swine flu vaccine & Pig, influenza, swine influenza, immunogen, disease, antigen, illness, etc. & Human flu, Influenza, Influenza virus, H2N3, H1N2, Influenza pandemic, flu vaccine, etc.  & Influenza viruses,  pig, influenza, H1N1, H1N2, human flu, world health organisation, Infectious disease, etc. \\ \hline
			128 & Godhra train attack & Train, coach, transport, fire, firing, passenger train,  human action, attack, etc. & Gujarat riots, Godhra train, Gujarat violence,Kar Sevak, Panchmahal district, etc.  & Sabarmati Express, train, coach, Godhra, Gujarat, Hindu, Ayodhya, Babri Masjid, etc.\\ \hline
			146 & Ram Janmabhoomi verdict & Ram, verdict, act, judgement, group action, human activity, judicial decision, etc. & Ram Janmabhoomi, Ayodhya, Babri Masjid, Baqi Tashqandi, Barabanki district, Bharatiya Janata Party, etc. & Ayodhya dispute, Hindu, Rama, avatar, Hindutva, Ramayana, Babri Masjid, Ayodhya, Allahabad High Court, etc. \\ \hline
		   174 & International economic slump & Global, economy, economic system, Market, drop-off, falloff, economic crisis, etc. & Economics, Demographic economics, Recession, Education economics, Markets, Public choice, Trade, etc. & Rescission, Macroeconomic, GDP, financial crisis, Stock market, Global recession, etc. \\ \hline
			
	\end{tabular}}
	%\end{center}
\end{table}	

\section{Conclusion}
\label{Conclusion}
This article presents a novel Wikipedia WordNet-based Query Expansion (WWQE) technique that considers the individual terms and phrases as the expansion terms. The proposed method employs a two-level strategy to select terms from WordNet. First, it fetches synsets of the initial query terms. Then, it extracts synsets of these synsets. In order to score the expansion terms on Wikipedia, we proposed a new weighting score named as an in-link score. We employed a tf-idf-based scoring system to assign a score to expansion terms extracted from WordNet. After assigning a score to individual query terms, we further re-weighted the selected expansion terms using the correlation score with respect to the entire query. The combination of the two data sources works well for extracting relevant expansion terms,  and the proposed QE technique performs well with these terms on several weighting models. It also yields better results when compared to the two methods individually and with the other related state-of-the-art methods. This article also investigates the retrieval performance of the proposed technique with varying numbers of expansion terms. The results on the basis of several evaluation metrics and popular weighting models on the FIRE dataset demonstrate the effectiveness of our proposed QE technique in the field of information retrieval. 

\section*{Acknowledgments}
Akshay Deepak has been awarded Young Faculty Research Fellowship (YFRF) of Visvesvaraya PhD Programme of Ministry of Electronics \& Information Technology, MeitY, Government of India. In this regard, he would like to acknowledge that this publication is an outcome of the R\&D work undertaken in the project under the Visvesvaraya PhD Scheme of Ministry of Electronics \& Information Technology, Government of India, being implemented by Digital India Corporation (formerly Media Lab Asia).

\section*{References}

\bibliographystyle{apa}
\bibliography{bibfile}

\end{document}